\documentclass[prl,aps,longbibliography,twocolumn,superscriptaddress]{revtex4}
\usepackage{latexsym}
\usepackage{amssymb}
\usepackage{amsmath}
\usepackage{graphicx}
\usepackage{dcolumn}
\usepackage{bm}
\usepackage{amsfonts}
\usepackage{hyperref}
\usepackage{color}
\usepackage{epsfig}
\usepackage{color}
\usepackage{natbib}

\begin{document}

\title{Many-Body Entanglement in Short-Range Interacting Fermi Gases for Metrology}

\author{Leonardo Lucchesi}
\affiliation{Dipartimento di Fisica “Enrico Fermi”, Universit\`a di Pisa, Largo B. Pontecorvo 3, I-56127 Pisa, Italy}
\author{Maria Luisa Chiofalo}
\affiliation{Dipartimento di Fisica “Enrico Fermi” and INFN, Universit\`a di Pisa, Largo B. Pontecorvo 3, I-56127 Pisa, Italy\\
JILA, University of Colorado, 440 UCB, Boulder, Colorado 80309, USA\\
Kavli Institute for Theoretical Physics, University of California, Santa Barbara, CA 93106-4030, USA}
\thanks{Corresponding author \href{mailto:maria.luisa.chiofalo@unipi.it}{maria.luisa.chiofalo@unipi.it}}

\begin{abstract}
We explore many-body entanglement in spinful Fermi gases with short-range interactions, for metrology purposes. We characterize the emerging quantum phases via Density-Matrix Renormalization Group simulations and quantify their entanglement content for metrological usability via the Quantum Fisher Information (QFI). Our study establishes a method, promoting the QFI to be an order parameter. Short-range interactions reveal to build up metrologically promising entanglement in the XY-ferromagnetic and cluster ordering, the cluster physics being unexplored so far.
\end{abstract}

\maketitle

Strongly-correlated systems are progressively becoming a paradigm for precision metrology, attracting broad interest~\cite{PezzeSmerziRMP}. Quantum gases represent a powerful platform to develop quantum measurement devices~\cite{yerev,giovrev}, bridging between engineering of quantum states of matter~\cite{SorensenMolmer} and progress in atom interferometry \cite{kasrev,TinoKasevich}. Atom interferometry has many sources of uncertainty, classifiable into device and statistics-driven causes~\cite{revatomint}. Accurate experimental schemes have blossomed, providing significant reduction of the former, now comparable or even lower than statistical error~\cite{revatomint, kaschu, mullerchu,gravconst,prlchiofalotino,kasnew}. Further precision improvements can be obtained by addressing the statistical uncertainty problem, in particular the quantum phase estimation~\cite{giov,pezzesmerzipest:TinoKasevich}.
A conceptual tool to reduce statistical uncertainty may come from entanglement, specifically quantum squeezing~\cite{manori,oberthaler,robertson,kitaueda}, where uncertainty in a selected observable can be reduced below the Heisenberg bound at expenses of a conjugate observable~\cite{walls}. Atomic spin squeezing has been implemented in numerous experimental setups, using interactions either collision-driven or light-mediated in optical cavities~\cite{vuleticsqueeze, Thompson2014,thomreyscience,ritschrev,manori}.\\ 
\begin{figure}
\centering\includegraphics[width=\columnwidth]{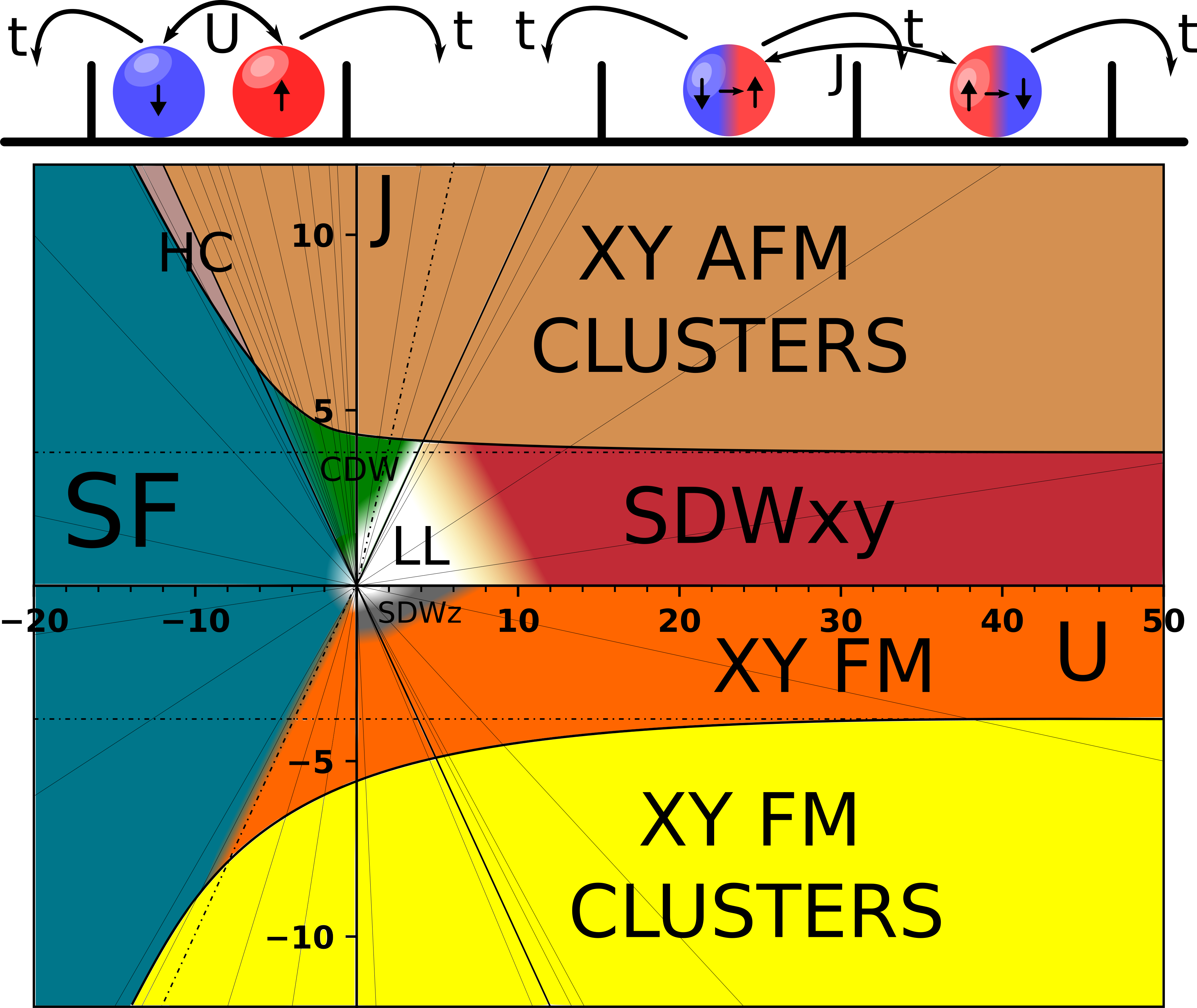}
\caption{(Color online) System concept. \textbf{Top}. The $tUJ$ Hamiltonian (\ref{eq:HtUJ}): $t$ drives the hopping, $U$ the on-site interaction and $J$ the spin-exchange coupling. \textbf{Bottom}. Qualitative phase diagram at quarter filling in the $U/t$-$J/t$ parameter space, including the following phases: Luttinger-Liquid (LL), Superfluid (SF), Charge-Density-Wave-like (CDW), Spin-Density-Wave ($\text{SDW}^{x,y,z}$), XY Ferromagnetic (XY-FM), Clusters with internal XY-FM or antiferromagnetic (AFM) spin ordering, and {\it hemmed} clusters (HC) (see text for descriptions).  Simulations have been performed along the solid lines. Thick solid straight lines: $|J/U|=1$. Thick curves: guidelines delimiting cluster phases. As $U\to+\infty$, $J_c\simeq 3.8$ separates AFM-like SDW$^{x,y}$ and XY-AFM Cluster phases, while $J_c\simeq -3.8$ separates XY-FM and XY-FM Clusters. Dot-dashed straight lines: studies from ~\cite{dziurzik} (tilted) and ~\cite{ogatasorella} (horizontal) (see text). We explore the metrological usability of these phases,
finding XY-FM and XY-FM cluster phases especially convenient (see text).}
\label{fig.1}
\end{figure}
Entanglement is a necessary but not sufficient condition for squeezing, its metrological usefulness being quantified via the Quantum Fisher Information (QFI) from Cram\'er-Rao bound for statistical estimation of variances ~\cite{Hyllus,pezzesmerzipest:TinoKasevich,PezzeSmerziRMP}.
Generation of useful entanglement is often performed by means of infinite-range interactions \cite{manori,gabpezzelepori,daley}, and can survive a power-law decay of the coupling~\cite{FossFeig}.
However, also many-body finite-range interactions can drive long-range correlations, reinforcing the need to account for particles indistinguishability~\cite{Eckert02} and making the quantification of entanglement an even more subtle issue, as witnessed by a timely debate~\cite{Zanardi,LoFranco,Lourenco,ZollerQFI} in both quantum information and many-body communities, also motivated by experimental observations in quantum gases~\cite{Kaufman}. The interesting question thus arises, whether short-range interactions can provide phases with useful entanglement content for metrology.\\
In this Letter, we tackle the problem from a conceptual perspective and investigate many-body entanglement via a minimal model able to reproduce the essential desirable features of a strongly-correlated quantum fluid with short-range interactions and motional degrees of freedom~\cite{salvisqueeze}. To this aim, we consider a system of $N$ fermionic atoms in two spin states within the $tUJ$ model~\cite{dziurzik}, correlated via nearest-neighbor coupling $J$ and on-site $U$, and in the presence of tunneling processes $t$.
We use Density-Matrix Renormalization Group (DMRG) simulations to characterize the system quantum phases and classify them by finding a quantitative correspondence between the QFI and the order parameters characterizing the quantum fluid, conveying two central messages. First, this idea acquires methodological significance, since QFI can be seen as an order parameter. Second, two particular ground states in a short-range interacting system result especially promising for metrological use, because of their QFI scaling with the number of atoms $N$. These phases correspond to an XY-ferromagnet and a cluster ordering, the latter being here identified and quantitatively analyzed in the whole $U$-$J$ phase diagram. Exploiting this metrological usability requires the devising of suited protocols~\cite{Nolan}, which we will discuss along with possible experimental realizations.\\
\noindent\textit{\textbf{The Fermionic $tUJ$ model- }}We consider an ensemble of fermions in two (real or pseudo)-spin states, moving in a one-dimensional (1D) geometry in the presence of a short-range interaction. We model the system as cartooned in top Fig.~\ref{fig.1}, according to the $tUJ$ Hamiltonian:
\begin{equation}
H=\sum_i \big[-t (c^\dagger_{i\sigma}c_{i+1 \sigma}+\text{h.c.})+U n_{i \uparrow} n_{i\downarrow}+J(s^+_is^-_{i+1}+\text{h.c.})\big].           \label{eq:HtUJ}
\end{equation}
Here, $c^{(\dagger)}_{j,\sigma}$ are destruction (creation) operators for fermions with spin $\sigma$ on site $j$, $n_j \equiv \sum_\sigma c^\dagger_{j\sigma}c_{j\sigma}$ is the number operator, and $s^{+(-)}_j\equiv c^\dagger_{j\uparrow(\downarrow)}c_{j\downarrow(\uparrow)}$ the spin raising (lowering) operators. The $t$-term mimics atomic motion via hopping. The $U$ and $J$ terms represent, respectively, the contact and nearest-neighbor parts of a same two-body interaction, from now on in $|t|=1$ units.\\
We explore the quantum phases of the $tUJ$ model by resorting to a DMRG method~\cite{white,schollwoeck,rossini}, as described in detail in the Supplemental Material~\cite{SoM}, which includes Refs.~\cite{legeza,andersson,friedelosc,Barndorff,Smerzi_Loschmidt,wineland,japaridze}.
\begin{figure}[htbp]
\centering\includegraphics[height=0.9\columnwidth]{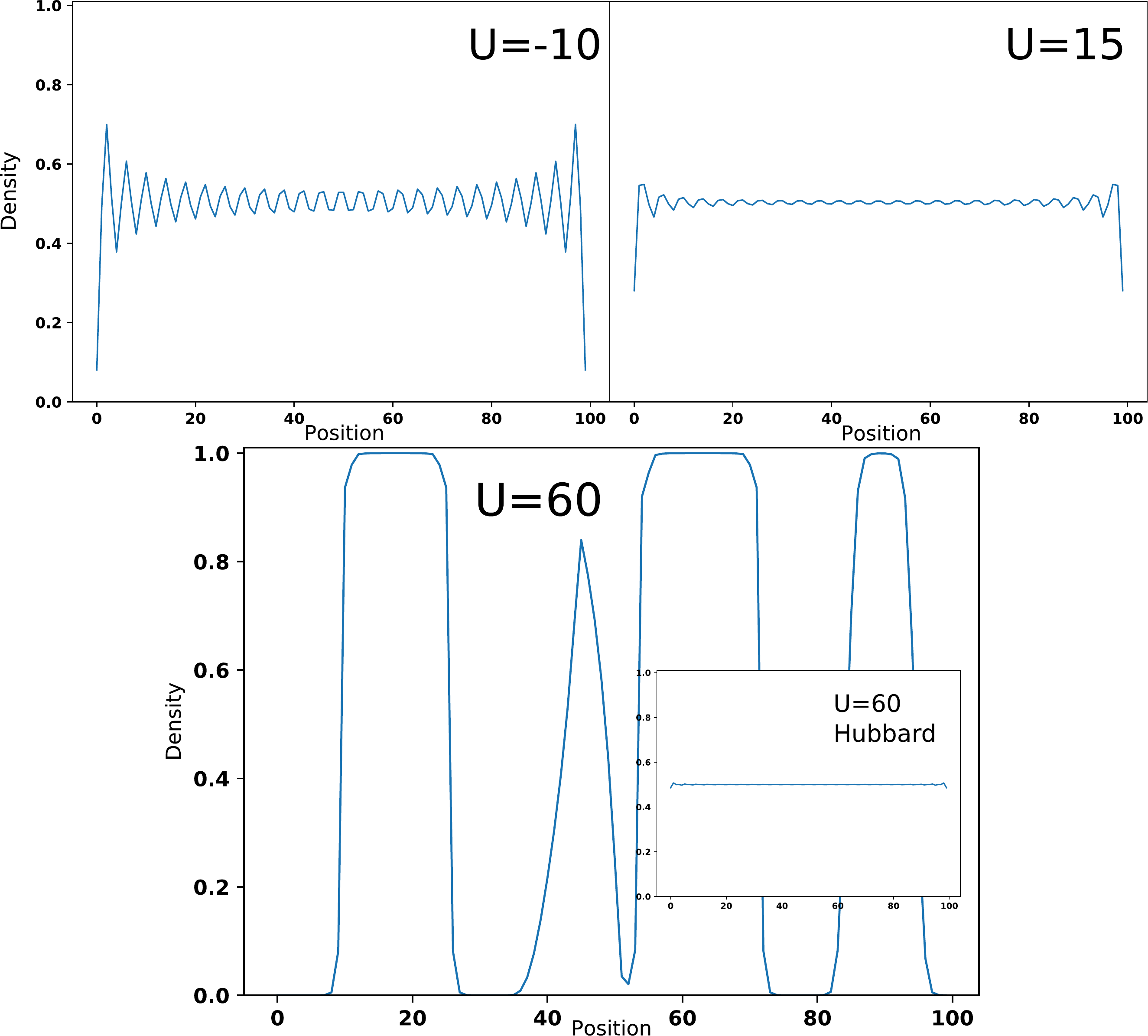}
\caption{Density profiles for $J=-0.1U$ and different $U$ values. Top: typical profiles in the SF and CDW (left), SDW and XY-FM phases (right). Friedel oscillations are present\cite{SoM}. Bottom: Density profile for $U=60$ with cluster formation. Inset: same profile with $J=0$.}
\label{fig.2}
\end{figure}
We probe different quantum correlation functions $\langle O^\dagger_i O_j \rangle$, with $O_k$ an operator acting on site $k$. We have considered Spin Density Waves (SDW) correlations with $O=s^{x,y,z}$, Charge Density Waves (CDW) with $O=n$, and superfluid pairing (SF) with $O=c_{\uparrow}c_{\downarrow}$. As we focus on the connection between the system quantum phases and their metrological usability, we only display results for $\nu=1/4$ filling, though results at $\nu=1/2$ are also discussed.\\ 
\textit{\textbf{Quantum phases- }}
The bottom Fig.~\ref{fig.1} displays the system quantum phases. We first discuss the phase diagram for $-\infty<U<+\infty$ and $|J|$-values below the solid thickest curves. Large and negative $U$ favor a SF phase in 1D sense with a large fraction of doubly-occupied sites~\cite{giamarchi}, while small $J$ couplings are ineffective without opposite-spins to pair. Moving towards $U\rightarrow0$, onsite pairs progressively become disfavored, and hopping start to dominate. As expected, this leads to CDW ordering for $J>0$ and SDW$^z$ for $J<0,U>0$, both characterized by a typical $2k_F$ oscillation in the correlation functions. Overall, the behavior around the origin is consistent with a smooth merging into a Luttinger-Liquid (LL) description. Larger and positive $U$ values drive instead a dominance of antiferromagnetic (AFM)-like ordering in the form of SDW$^{x,y}$ oscillating correlation functions for $J>0$. For $J<0$, a positive and non-oscillatory power-law behavior sets in, along with suppression of spin-$z$ correlations, while the spin-$x,y$ expectation values on each site are solid zeros. We call this XY-Ferromagnetic (XY-FM) phase in 1D sense, the power-law decay being the longest range ordering possible~\cite{giamarchi}. All this suggests the many-body ground state to be fully symmetric in the xy pseudospin plane, as dictated by the symmetry of the Hamiltonian. The SDW$^{x,y}$ and XY-FM phases can be understood noticing that $+J\sum_i(s^+_is^-_{i+1}+\text{h.c.})$ can be cast as $\sim s^x_is^x_{i+1}+s^y_is^y_{i+1}$, so that spin-exchange coupling favors spin (anti-)alignment in the $x,y$ plane.\\
We remark that a similar $tUJ$ model has been investigated by Dziurzik, Japaridze et al. \cite{dziurzik} in the context of high-temperature superconductivity via bosonization and DMRG techniques, exploring the $J,U$ space at different fillings.
While we find good agreement on the phases nature and boundaries discussed so far (tilted dot-dashed 
lines in Fig.~\ref{fig.1}~\cite{dziurzik}), our analysis provides qualitative and quantitative evidence of a new phase. In this phase, particles clusterize, i.e. form regions with unit density surrounded by zero density. Inside the clusters, spins are strongly aligned (FM) or antialigned (AFM) in their $x,y$-components. In Fig.~\ref{fig.1} these are the XY-FM and XY-AFM cluster phases, emerging for $J<0$ and $J>0$, respectively above and below a $U$-dependent threshold $J_c$. We now investigate the nature of these phases, turning our attention to the density profiles displayed in Fig.~\ref{fig.2} for the illustrative value $J/U=-0.1$~\cite{SoM}. 

While for $U<0$ and $U\lesssim 3$ values (top panel), the density profiles show the usual Friedel oscillations around average density~\cite{SoM}, for $U\gtrsim 38$ we encounter the typical situation depicted in the lower panel. The system's bulk ceases to be translationally invariant, and fermions form clusters of singly occupied sites. Simultaneously, very strong spin-$x$ correlations arise among particles inside clusters~\cite{SoM}. A similar simulation for the Hubbard model with $J=0$  shows no trace of this phase (inset), leading us to infer that the cluster phase be driven by the dominance of the local nearest-neighbor (FM and AFM) $xy$ coupling over the delocalizing hopping term. We assess the robustness of this phase by performing a number of runs against variations of simulation parameters. Though clusters positions and number are seen to change in sensible manner, their qualitative behavior persists as detailed in~\cite{SoM}. In essence, with our DMRG algorithm, single clusters more likely form at relatively small system sizes ($L\lesssim40$), and moving clusters may merge under larger numbers of finite-size algorithm iterations. We infer that the variability of the clusters positions be due to the vanishing energetic cost of moving around one of them in the surrounding free space.\\
In fact, we found traces of this state in studies of the tJ model performed via exact diagonalization~\cite{ogatasorella}, yielding $J_c=3.22$, and via DMRG, resulting in $J_c\simeq 3.15$~\cite{manmanadmrg}. From our density profiles, we infer that the cluster phase appears at $U_c\simeq 38$, i.e. - given $J/U=-0.1$ - $ J_c\simeq-3.8$. We infer that this phase transition is driven by the same physical mechanism as in~\cite{ogatasorella}, but with a critical $J_c$ modified by the onsite $U$. In fact, their no-double occupancy setting can be viewed as our $U\to+\infty$ limit, where we find $|J_c|\simeq 3.8$. For large $U<0$, the boundary is instead located on the lines $|J/U|\sim \pm 0.85$. As one would expect $|J/U|=\pm1$, the observed modified value could be due to super-exchange. 
In the $-1<J/U<-0.85$ gap, we observe peculiar clusters characterized by double-occupancy at the density edges, which we name hemmed clusters (HC)~\cite{SoM}. This is not the case in the symmetric region with $J/U>0$.\\
\textit{\textbf{Quantum Fisher Information (QFI)- }}
Having characterized our quantum phases, we can now turn to measure their degree of many-body entanglement via the Quantum Fisher Information $F$, and test the system's metrological usability. The quantum Cram\'er-Rao lower bound \cite{pezzesmerzipest:TinoKasevich} on an estimator variance is given by $(\Delta \theta)^2=1/F[\rho,\hat S]$.
The QFI depends in a complicated way on both the system's initial state and the transformation performed by the physical phenomenon to be measured, but it considerably simplifies for a pure state undergoing a unitary transformation $\exp{(i\theta S_{\vec{a}})}$, becoming $F[\psi,\hat S_{\vec{a}}]=4(\Delta S_{\vec{a}})_{\psi}^2$ \cite{pezzesmerzipest:TinoKasevich}.
Here $S_{\vec{a}}\equiv a_\alpha S^\alpha$ is a linear combination of global (pseudo-)spin operators \cite{pezzesmerzipest:TinoKasevich}. $F[\psi,\hat S_{\vec{a}}]$ fixes a criterion for evaluating the metrological usability of a quantum state, here the ground state of the many-fermion system. It is known that for a N-body uncorrelated product state, $F\sim N$ corresponds to the shot-noise limit \cite{pezzesmerzipest:TinoKasevich}. For possibly good metrological usability then, the QFI needs to scale as $N^\gamma<$,with $1<\gamma<2$ limited by the Heisenberg principle~\cite{giov}.\\
\begin{figure}
\centering\includegraphics[width=0.9\columnwidth]{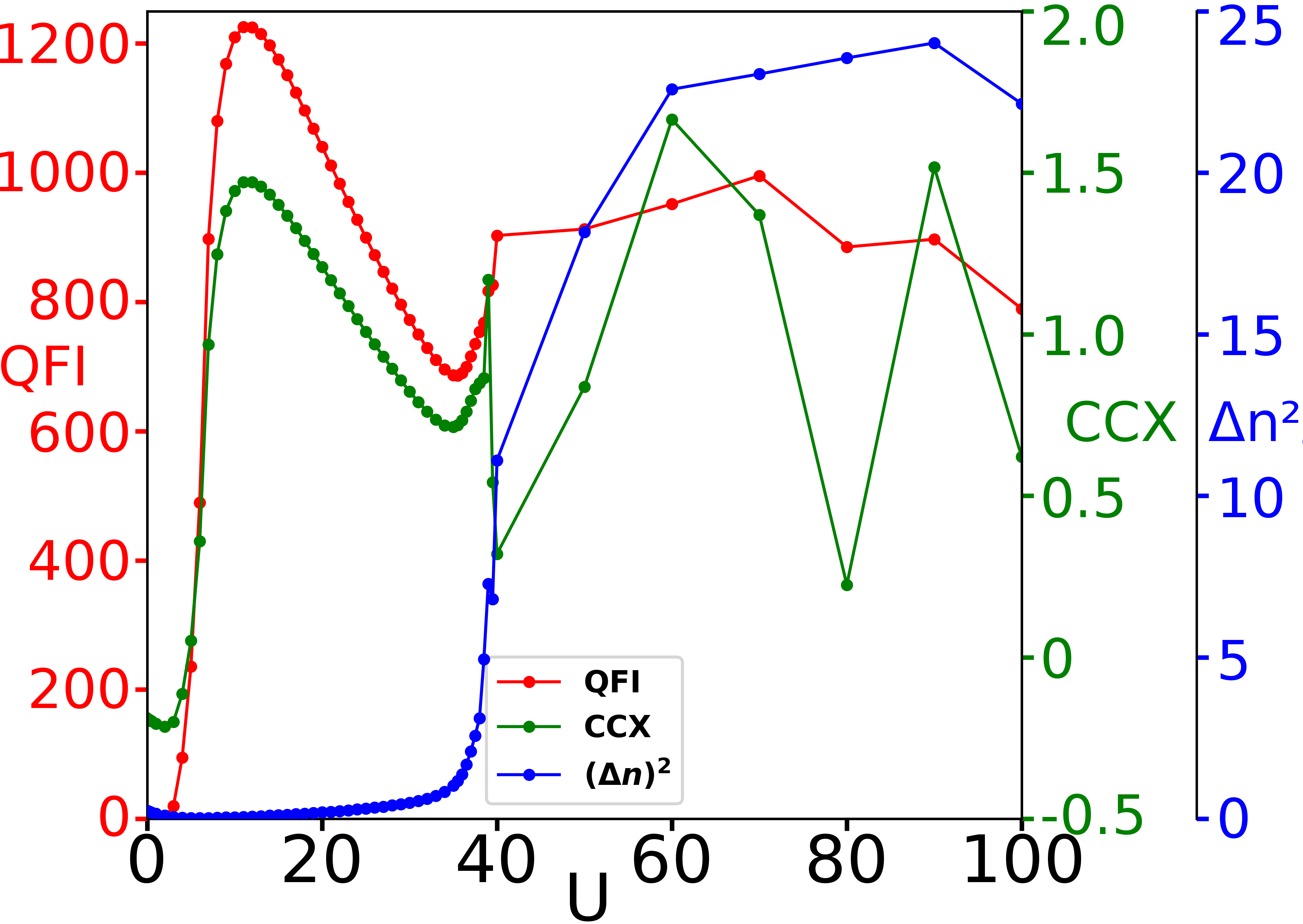}
\caption{Quantum fermionic correlated phases and metrological usability in a single-shot phase diagram. The QFI (red) vs. $U$ gets along the order parameters $CC_x(0)$ (green) and $\sum_i (\Delta n^2)_i$ (blue) describing the building up of XY-FM and Cluster correlations, respectively (see text).}
\label{fig.3}
\end{figure}
\textit{\textbf{Results on QFI- }}
We now quantify these expectations by computing the QFI across the phase diagram and comparing it with the quantum phases order parameters. In all computations we select the spin axis which offers the largest QFI value from the the angular momentum covariance matrix $\text{Cov}_{ab}=\sum_{i,j}\langle s^a_i s^b_j\rangle$~\cite{manori}, always obtaining the $x$-axis as non-granted outcome.
A simple reasoning would lead us to infer that the QFI on SDW or SF states would return a tiny value as compared even to shot-noise QFI$\sim N$. In fact, 
the oscillating spin-$x$ correlations between different sites would add up to zero in the $SDW$ and vanish for each doubly occupied site of the $SF$ state. This view corresponds to our numerical findings. The QFI results to be large only in the XY-FM and XY-FM cluster phases. For a quantitative comparison, we now define the corresponding order parameters. For the XY-FM phase, this is taken to be the area $CC_x(0)$ of the normalized $k=0$ peak in the Fourier transform of the spin-$x$ correlation function $C_x(i-j)$. For the Clusters phase, it is the normalized density variance $L^{-1}\sum_i (\Delta n^2)_i$.\\
Since we are originally interested in systems where $J$ and $U$ are effectively caused by the same term, we run simulations at fixed $J/U$ while varying $U$ to cross all possible phases. The results for the QFI (red points and curve), XY-FM (green points and curve), and Cluster (blue points and curve) order parameters are collected within one single graph in Fig.~\ref{fig.3}, one central result of the present work. We see that the QFI shows a steep change in correspondence of the quantum phase transition to spin-$x$ ordering, the QFI and $CC_x(0)$ curves getting quite closely along with varying $U$. In fact, one may use the QFI to infer the occurrence of the two quantum phase transitions around $U\sim 4$ and $U\sim38$. The correspondence between QFI and order parameters is quantitative for the XY-FM phase. The fact that particles in different clusters are uncorrelated makes the comparison qualitative for the Cluster phases at this stage. A quantitative treatment is recovered via the QFI scaling analysis below, generalizing this central message to different $J/U$ values in the phase diagram.\\
\begin{figure}
\centering
\includegraphics[width=\columnwidth]{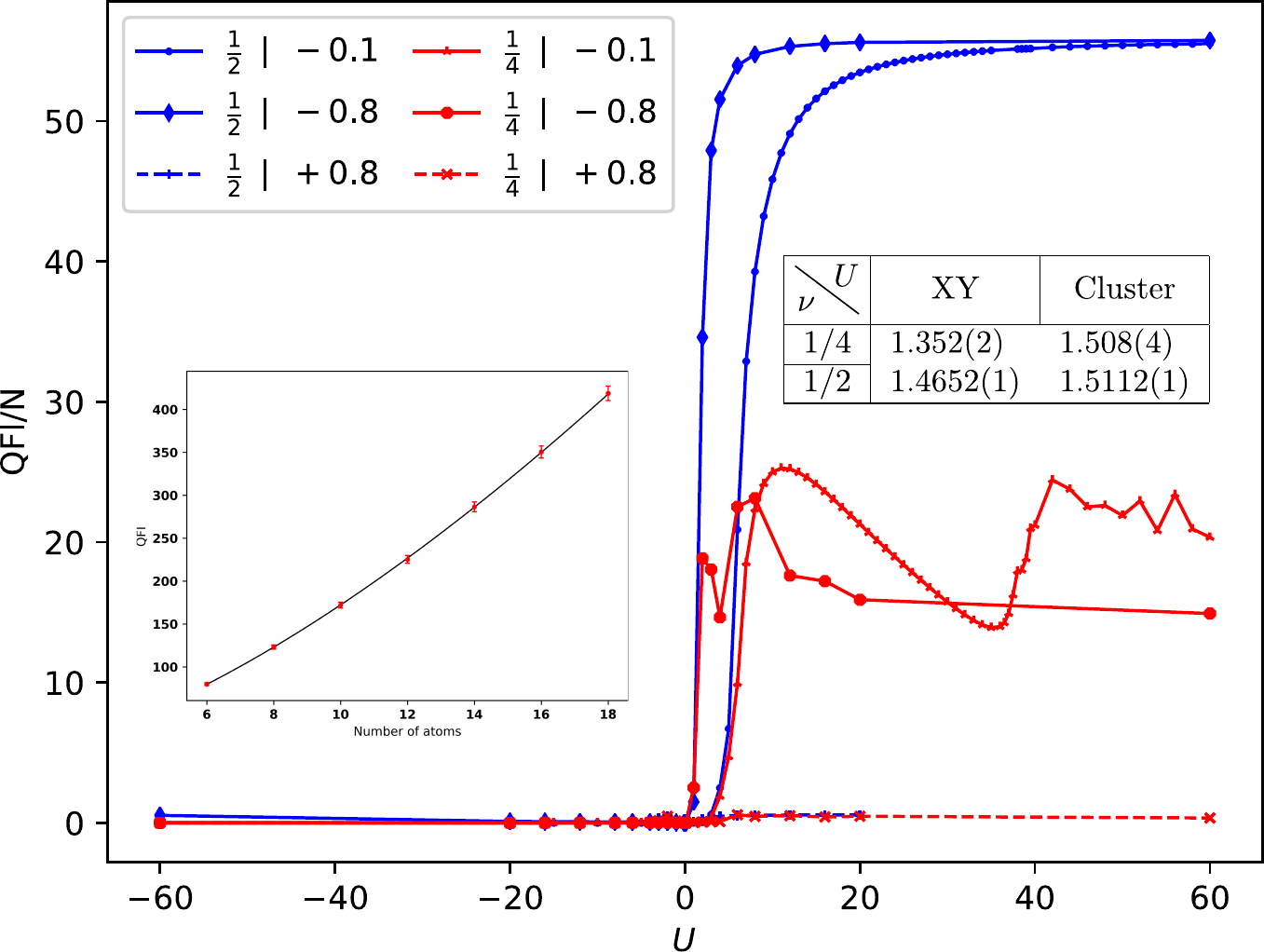}
\caption{QFI density QFI/N ($N$ number of atoms) vs. $U$ for $\nu=1/4$ (red) and $1/2$ (blue), and different $J/U$ (see legend). Table: exponents fitted from QFI$=kN^\gamma$ for $\nu=1/2$, $1/4$ at  $J/U=-0.1$. $U$ values are chosen to correspond to the QFI maximum in the XY-FM phase ($U=11$) and in the large-$U$ limit for the Cluster phase ($U=60$), at $\nu=1/4$. Inset figure: example of QFI scaling for $\nu=1/4$, $U=60$, and $J/U=-0.1$.
}
\label{fig.4}
\end{figure}
In particular, we now study the dependence of QFI on $J/U$ and filling, and assess the degree of metrological usability from the QFI scaling with the particle number $N=2\nu L$~\cite{PezzeSmerziPRL}.
We display in Fig.~\ref{fig.4} the QFI density QFI/N at two commensurate fillings, $1/4$ (red) and $1/2$ (blue). As anticipated, the QFI vanishes for $U<0$ and $J/U=+0.8$, where XY-FM and cluster phases are absent. At $\nu=1/2$, the QFI density is larger and, unlike $\nu=1/4$, smooth since the whole system is in the form of a single cluster. For both fillings, larger (negative) values of $J$ favor cluster formation and steeper QFI rise.\\
We study the $N$-scaling with special care at $\nu=1/4$, where several uncorrelated clusters may form at large $U>0$. Thus, we keep relatively small system sizes ($L<40$) to have one single cluster\cite{SoM}. For both fillings, we fit the QFI dependence on $N$ with QFI$=kN^\gamma$, as illustrated in the inset.
The table reports $\gamma$ for $U=11$, corresponding to the QFI maximum in the XY-FM phase, and the large-$U$ limit for the Cluster phase at $\nu=1/4$. We see that half-filling shows better scaling outside the cluster region. Inside it, the scalings at $\nu=1/4$ and $1/2$ are compatible within error.\\
\textit{\textbf{Metrology implementations.}} The QFI scaling is promising, but a real use of these reduced-quantum uncertainty states injected in an interferometric sequence requires suited protocols. The XY-FM and Cluster phases represent non-Gaussian states with a Wigner distribution located around the equator in the Bloch sphere~\cite{SoM} and $\langle S_{x,y,x}\rangle=0$, so that the signal cannot be encoded in a mean spin direction. This unconventional situation reminds the one experimentally investigated in~\cite{Lucke} for Twin-Fock states with the method proposed in~\cite{Kim}. Adopting a similar strategy, one might operate a rotation by angle $\theta$ about an axis in the xy-plane, and consider the lower bound $\mathcal{F}\geq |d\langle S_z^2\rangle/d\theta|^2/(\Delta S_z^2)^2$ for the classical Fisher information, leading to the uncertainty $\Delta\theta\geq (\sqrt{\mathcal{F} n)}$ after $n$ measurements. In essence, the signal would be related to the second moment of $S_z$ instead than the first one, and the noise to the fourth instead than the second. Eventually, optimization with respect to $\theta$ is to be performed. Signal extraction and optimization can be operated after sampling the full probability distribution or the second and fourth $S_z$ momenta~\cite{Strobel} in a time-dependent simulation of the interferometric sequence.\\ 
\textit{\textbf{Conclusions- }} Our study conveys two unforeseen messages. First, short-range interactions are able to build metrologically useful entanglement in a many-fermions system. This is demonstrated by a large degree of Quantum Fisher Information, accompanied by interesting scaling with the number of 
particles. The best performing phase is indeed the cluster one, driven by the $J$ coupling,  which in our study models the short-range interactions. Second, our results imply that the QFI represent a powerful tool to characterize the phases of the quantum fluid, acting as an order parameter.\\ 
Implementations in ultracold gases platforms may in include currently realized systems of dipolar fermions in optical lattices~\cite{Ferlaino} and suitably engineered versions of Fermi-Hubbard setups~\cite{greiner}, in both cases after further reduction of dimensionality to 1D. Finally, a microscopic origin of this $tUJ$ model can be provided by a photon-mediated effective interaction among fermions in an optical cavity~\cite{elviaprof}, leading to a spin-squeezing-like Hamiltonian~\cite{manori}. Multimode optical cavities~\cite{lev} may bring in the short-range environment, though a realistic probe requires detailed modeling to include unavoidable
dissipation processes~\cite{keeling}. Single-particle decoherence could be suppressed in the presence of a spin gap, as in the cluster phase~\cite{dziurzik}. While one might expect superradiance-enhanced decoherence still be an issue, one might ask whether delocalization in \eqref{eq:HtUJ} and the xy-symmetric structure of the ground state might be exploited to limit the effect. We are currently working along this direction, via an actual time-dependent simulation of the open system~\cite{keeling}.\\

\begin{acknowledgments}
We thank Davide Rossini for valuable support in using his DMRG code. We are grateful to Augusto Smerzi for enlightening discussions on actual protocols. We thank Luca Lepori and Luca Pezz{\'e} for useful discussions, and Benjamin Lev, Jonathan Keeling, and Andrew Daley for collaborative work on optical-cavities implementations. M.L.C. thanks JILA for fruitful and warm hospitality during the visiting fellowship, when part of this work has been carried out, and in particular Ana Maria Rey and Murray Holland for enriching discussions. M.L.C.  would like to thank KITP for hospitality during the program on Open Quantum Systems, when discussions have revealed to be useful also to envisage possible follow-ups of this work. This research was supported in part by the National Science Foundation under Grant No. NSF PHY-1748958. We acknowledge the MAGIA-Advanced project for support and the Goldrake HPC team.
\end{acknowledgments}


%

\newpage

\begin{center}
\huge Supplemental Material for: \\ Many-body Entanglement of Short-Range Interacting Fermi Gases for Metrology
\end{center}


\section{DMRG method and checks.}
DMRG is a powerful numerical method that has been extensively employed so far to investigate spin Hamiltonians and both spinless and spinful itinerant systems~\cite{white,schollwoeck}.\\
We summarize below the relevant simulation steps in the algorithm, which incrementally builds the global system Hamiltonian from a small solvable system. 
At first, all the terms for a new site are added to the Hamiltonian. Then, the ground state is found by using a fast algorithm (Davidson). The ground state is then represented by a truncated Hilbert space, obtained by keeping only the first $m$ vectors with the largest eigenvalues in the basis that diagonalizes its corresponding density matrix. After projecting each operator on the truncated Hilbert space, the system is ready for the next step.\\
When the truncation happens, the Hilbert space spun by the $m$ vectors becomes the new Hilbert space of the whole system, implying that some of the fine structure of the system is neglected. The quantity $m$ is thus an important parameter of the simulation and must be  carefully chosen, in order to balance between precision and computational speed. Due to its peculiar constructive procedure, DMRG is best suited for short-range interacting (1D) systems as one needs to include all the interaction terms from the new site at each step. A long-range interaction would mean to add lots of new terms and to keep many single sites for each step, as the interaction Hamiltonian would couple them separately and not as a bulk.
When the system reaches the desired length $L$, the DMRG procedure can be modified to preserve the number of sites.
For the present problem we have adapted the code provided by Rossini et al.~\cite{rossini}, which implements the \textit{finite size} DMRG algorithm, a procedure which optimizes the representation using two neighboring sites per step, and improves the thermalization procedure already included in the standard DMRG algorithm \cite{schollwoeck}.
This finite-size step is performed sequentially, sweeping all the 1D chain back and forth $N_s$ times, with $N_s$ a parameter of the simulation. This part has good convergence properties and usually just a few sweeps ($N_s\sim3$) are enough \cite{rossini}, though we use up to $N_s=10$ sweeps for benchmarking purposes and to investigate merging in the cluster phase.
In order to deal with the fermionic system, we finally have implemented in the code the Jordan-Wigner transformation.\\ 
The simulational parameters are the number of vectors spanning the truncated Hilbert space $m$, the system length $L$ and the number of sweeps $N_s$. As no fully standard procedure is given to tailor them \cite{schollwoeck,legeza,andersson,rossini}, we performed a numerical analysis to fix the best values satisfying the following criteria: (i) a proper bulk with reduced Friedel oscillations, a phenomenon due to the open boundary conditions implied in the DMRG procedure \cite{friedelosc}; (ii) a better representation of the long-range correlations while avoiding a numerical correlation-length effect \cite{andersson}; and (iii) an optimal trade-off between the above goals for fast convergence.

\begin{figure}[bt]
\includegraphics[width=0.95\columnwidth]
{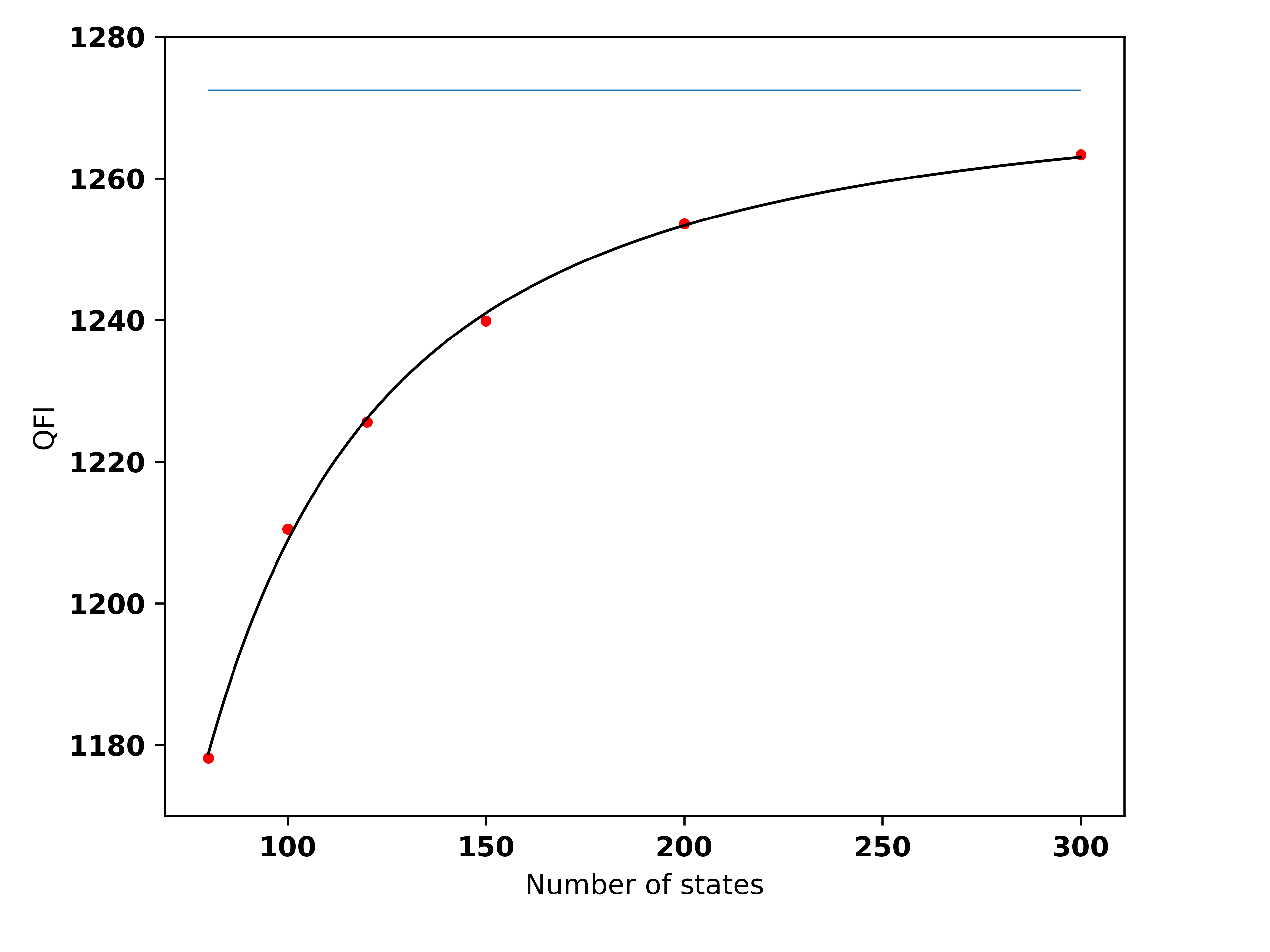}
\caption{Numerical checks. Illustrative example of the adopted extrapolation procedure. Convergence of the QFI with the number of states $m$ in the truncated basis. The fit has been performed by using a power law $F_\infty - k m^{-\gamma}$, obtaining $F_\infty=1272$. Thus, with $F_{300}=1263$ we can estimate the truncation error for $m=300$ to be $\sim 1 \%$ in this regime and filling. The checks suggest no qualitative differences between different $m$ cases.}
\label{smfig1}
\end{figure}
Typical runs are performed with $L=100$, $m=300$, and $N_s=3$, though simulations with different values have been carried out,  especially for checking purposes on energy and  order-parameter scaling with $m$. We have then extrapolated the $m\rightarrow\infty$ value by performing a fit for any quantity employed in this work as illustrated in Fig. \ref{smfig1}. The values provided for the different quantities are indeed the result of this extrapolation procedure. The error on the fit parameter was taken as the error on the quantity. As the number $L$ of sites is related to the number $N$ of atoms via the filling, the scaling in $L$ has a physical meaning, as discussed in the main text.\\

\subsection{Clusters and sweeps}
As we encountered the previously unexplored cluster phase, we want to exclude any numerical influence in the formation of clusters. A numerical check that is also physically meaningful amounts to test for different numbers of finite-size algorithm iterations (sweeps).\\
\begin{figure}
\includegraphics[width=0.95\columnwidth]
{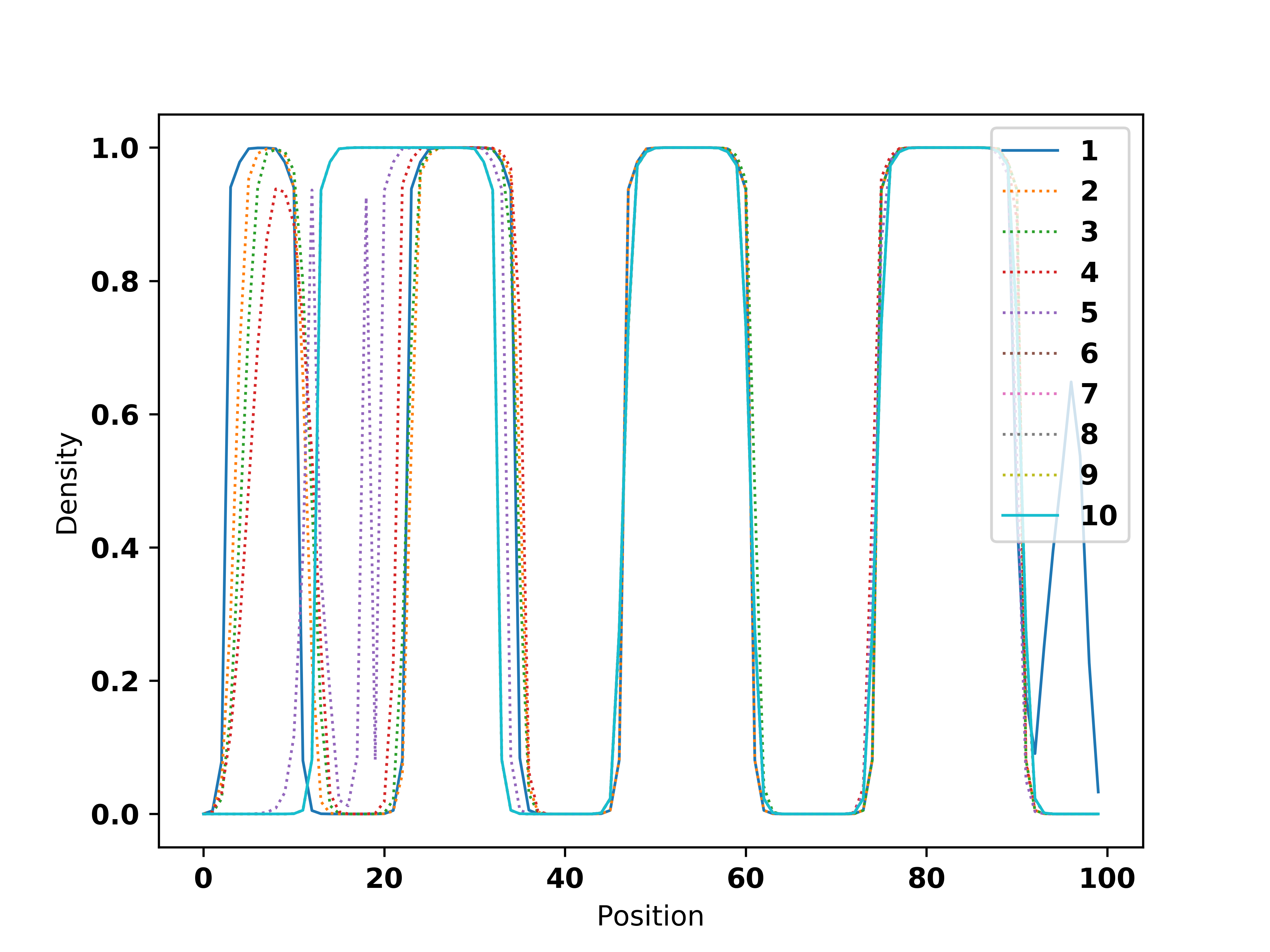}
\caption{ Clusters and sweeps. Variations of clusters positions with number of finite-size DMRG iterations (sweeps). We can directly observe the evolution of cluster positions with the number of sweeps. Two observations can be made: clusters are able to merge, and sweeps can cause this merging.}
\label{smfig2}
\end{figure}
As showed in Fig.\ref{smfig2}, sweeps can cause cluster motion and merging. This is due to the fact that the ground state is strongly degenerate, preventing extraction of definite information on clusters positions. DMRG is not the best suited method for studying this peculiar phenomenon, as it performs a series of \textit{local} (2 sites wide) optimizations in order to find the ground state. In principle this would prevent motion of clusters extending over more than 2 sites, since moving them around would require to break them up, with large energy cost. In practice, we find that clusters can move in worm-like steps, by extruding and pulling back small parts. When two extruded parts overlap, the merging can begin. This process is energetically favored because of the presence of a surface energy, that can be traced back to the spins in cluster edges interacting with only one spin in the inner cluster. In fact, if a spin at the cluster edge were to interact with \textit{two} spins, it would further lower the total energy.
\subsection{About the use of DMRG in QFI scaling vs. $N$ in the XY-FM and Cluster-phases}
For $U=11$ in the XY-FM phase, we computed the QFI for $L$ ranging between 40 and 200 with $m=300$, the values and the uncertainties being chosen following the convergence procedure already discussed.
Computing the QFI scaling for the cluster phase at $\nu=1/4$ (when multiple clusters may form) requires additional care, due to the issue of cluster positions and sizes. Separated clusters are uncorrelated, giving smaller QFI values than a single cluster configuration. However, we noticed that for relatively small sizes ($L< 40$), the system prefers to form one single cluster. This size limit depends on the value of $U$, so we get around the algorithm problem by performing a series of simulation runs with fixed $U=60$ and $m=300$ and variable $L<40$. The resulting cluster phase is at this point stable with respect to boundary effects, due to easier matching of the zero density at the box boundaries (open boundary conditions). No such a problem occurs at half filling, since only one cluster always forms, irrespective of the parameters values in the cluster region.

\section{Quantum Fisher Information}

   In the most general description of quantum mechanics, we describe the state of a system by a matrix $\hat \rho$, with complex coefficients, that embodies a classical statistic description of our system's state occupation. The properties of this (density) matrix ensure a classically significant distribution probability on a determined set of states. This matrix can (like a state) depend on a parameter $\theta$, the one we would like to estimate.\\ 
            In this picture, measurements are described by defining a set of projectors $\hat E(\epsilon)$, whose corresponding Hilbert subspaces are the eigenspaces relative to each eigenvalue $\epsilon$ of the operator representing the physical quantity we want to measure.\\
            The probability of observing $\epsilon$ considering both classical and quantum probability contributions is given by
            \begin{equation}\label{expvalrhotheta}
                P(\epsilon|\theta)=\text{Tr}\Big[\hat E(\epsilon) \hat \rho(\theta)\Big],
            \end{equation}
            that has a clear meaning if we see it back in the pure states picture.\\
            If the state is pure, $\hat \rho=|\psi\rangle\langle\psi|$, and the projector is defined as $\hat E(\epsilon)=|\phi(\epsilon)\rangle\langle\phi(\epsilon)|$, so the former expression yields again $|\langle\phi(\epsilon)|\psi\rangle|^2$.
            This is also called \textit{likelihood} and it is the conditional probability of obtaining $\epsilon$ being in the state $\psi(\theta)$, thus having $\theta$ as the parameter value.\\
            Within this framework, we can perform a statistical estimation of the parameter $\theta$ by trying to reconstruct the probability distribution of $\epsilon$.\\
            This is made by defining an estimator $\Theta(\theta)$, which is subject to a lower bound on uncertainty.
         \subsection[Quantum Cramer-Rao bound: definition of the QFI]{Quantum Cramer-Rao bound: definition of the Quantum Fisher Information}
            The Cram\'er-Rao inequality sets a lower bound on the uncertainty of an estimator $\Theta$. We can recall its relations as:
            \begin{equation*}
                 (\Delta \Theta)^2_\theta \ge {\big({{ {\partial \langle \Theta \rangle_\theta} \over { \partial \theta}}\big)^2 }\over {I(\theta)}}
               \end{equation*}
               with the classical Fisher information (CFI) defined as 
               \begin{equation*}
                  \mathcal{F}(\theta) \equiv \bigg\langle \bigg(\frac{\partial L(\epsilon\vert \theta) }{ \partial \theta}\bigg)^2 \bigg\rangle_\theta = \sum_{\epsilon} \frac{1}{P(\epsilon | \theta)} \bigg( \frac{\partial P(\epsilon | \theta)}{\partial \theta} \bigg)^2.
            \end{equation*}
            The proof of this inequality is usually part of a standard statistics course, and can be found in \cite{pezzesmerzipest:TinoKasevich}.
            The quantum mechanical result for the Fisher information, obtained by combining its definition with \eqref{expvalrhotheta}, depends on our choice of the set of projectors $\hat E(\epsilon)$ and therefore on the observable we are using for estimating our $\theta$. We can then define a new quantity, the \textit{quantum Fisher information}
            \begin{equation}\label{defQFI}
                F\big[ \hat \rho (\theta) \big] \equiv  \max_{ \{ \hat{E}(\epsilon) \} } \mathcal{F}\big[ \hat \rho (\theta), \{ \hat E(\epsilon) \} \big],
            \end{equation}
            that is by definition the biggest chunk of information that we can obtain from our ensemble of systems, or else the smallest variance for our estimation. We shall then link our definition to actual physical quantities, because computing the Quantum Fisher Information for each possible observable might be a bit tedious.\\
            We would like to have something equivalent to the classical expression, that can be written as
            \begin{equation}
                \mathcal{F}(\theta)\equiv\sum_{\epsilon} \frac{1}{P(\epsilon | \theta)} \bigg( \frac{\partial P(\epsilon | \theta)}{\partial \theta} \bigg)^2=\sum_{\epsilon}P(\epsilon | \theta)\bigg( \frac{\partial \ln P(\epsilon | \theta)}{\partial \theta} \bigg)^2,
            \end{equation}
            such that the Fisher information can be cast as the expectation value of the square of the quantity that we have called  \textit{logarithmic derivative}.\\
            It can be proved that the Quantum Fisher Information can be written as \cite{pezzesmerzipest:TinoKasevich}
            \begin{equation}
                F\big[ \hat \rho (\theta) \big] = \text{Tr}\big[ \hat{\rho}(\theta) \hat L^2_{\theta} \big],
            \end{equation}
            where we define the Symmetric Logarithmic Derivative (SLD) $\hat L_{\theta}$ as the Hermitian operator which is  solution of the equation:            
            \begin{equation}\label{defsld}
                \frac{ \partial \hat{\rho}(\theta)}{\partial \theta} = \frac{ \hat{\rho}(\theta) \hat L_{\theta} + \hat L_{\theta} \hat{\rho}(\theta)}{2}.
            \end{equation}
           This is an expectation value of an operator, much simpler than finding the maximum value of all observables. Still, finding $\hat L_{\theta} $ is not an easy task because of the operator-valued differential equation that needs to be solved in order to find it. In the following section, we will find what can be done only by using its definition.
         \subsection{Quantum Fisher Information: exploring the definition}
         Here we recall a set of analytical expressions, useful for computing the QFI in many real applications.
            \paragraph{Pure states.}
                The definition of the Symmetric Logarithmic Derivative (from now on SLD) greatly simplifies when computing it on a pure state. The density matrix of a pure state is a projector, so $\rho^2(\theta)=\rho(\theta)$. Then its derivative can be written as
                \begin{equation}
                    \partial_\theta \rho = \partial_\theta \rho^2= (\partial_\theta \rho) \rho +  \rho (\partial_\theta \rho),
                \end{equation}
                allowing us to identify $\hat L_{\theta}=2\partial_\theta \rho$ with the definition \eqref{defsld}.\\
                We can rewrite $\rho=|\psi\rangle\langle\psi|$, so that the QFI can be cast in the form
                \begin{equation}\label{qfipurstat}
                    F\big[ \vert \psi (\theta) \rangle \big]= 4 \big( \langle\partial_\theta \psi|\partial_\theta \psi\rangle - \vert \langle\partial_\theta \psi|\psi \rangle \vert^2 \big).
                \end{equation}
            \paragraph{Mixed states.}
                The best way to handle this case is to write everything using the (always existing) eigenbasis of the density matrix $\rho$.\\ 
                Let us say that $\rho(\theta)=\sum_k p_k |k\rangle\langle k| $ with $p_k$ being the weight of the state $|k\rangle$. Thus, we can rewrite the definition of the QFI containing the logarithmic derivative using this basis:
                \begin{equation}
                    F\big[ \hat \rho (\theta)\big] = \sum_{k,k'} p_k \big\vert \langle k \vert \hat L_{\theta} \vert k' \rangle \big\vert^2 = \sum_{k,k'} \frac{p_k+p_{k'}}{2} \big\vert \langle k \vert \hat L_{\theta} \vert k' \rangle \big\vert^2.
                \end{equation}
                We can find the matrix elements of $\hat L_{\theta}$ in this basis by writing its definition in this basis:
                \begin{equation*}
                \begin{array}{rl}
                    \langle k | \partial_{\theta} \rho |k'\rangle & =\frac{1}{2}\Big( \langle k | \rho \hat L_{\theta} |k'\rangle + \langle k |\hat L_{\theta}\rho|k'\rangle\Big)\\
                    & =\frac{1}{2}\Big(p_k\langle k |\hat L_{\theta} |k'\rangle + p_{k'}\langle k |\hat L_{\theta}|k'\rangle\Big)\implies\\
                    \implies\langle k | \hat L_{\theta} | k' \rangle& =2 \frac{\big{\langle k \vert \partial_{\theta}\, \rho(\theta)\vert  k'\rangle}}{\big{(p_ k + p_k')}}.
                \end{array}
                \end{equation*}
                Therefore, we can rewrite the former equation as
                \begin{equation}
                    F\big[ \hat \rho (\theta) \big]= \sum_{k,k'}\frac{2}{p_k + p_{k'}} \big|  \langle k| \partial_\theta  \rho (\theta) |k'\rangle \big|^2.
                \end{equation}
                We shall write $\partial_{\theta}\rho$ in the eigenbasis
                \begin{equation*}
                \begin{array}{cl}
                    \partial_{\theta} \rho(\theta)& = \sum_k \big(\partial_{\theta} p_k \big) |k\rangle\langle k| + \\
                    &\sum_k p_k |\partial_\theta k\rangle \langle k| + \sum_k p_k |k\rangle \langle\partial_\theta k|\\
                    & \\
                    \langle k| \partial_\theta  \rho (\theta) |k'\rangle& = (\partial_{\theta} p_k) \delta_{k,k'} +(p_k - p_{k'}) \langle\partial_\theta k| k' \rangle .
                \end{array}
                \end{equation*}
                The final expression for the QFI reads:
                \begin{equation}
                    F\big[ \hat \rho (\theta) \big]= \sum_k \frac{(\partial_\theta p_k)^2}{p_k}+ 2 \sum_{k,k'} \frac{(p_k - p_{k'})^2}{p_k + p_{k'}}\big\vert \langle\partial_\theta k| k' \rangle \big\vert^2.
                \end{equation}
                We can even write the SLD directly in this basis in the form:
                \begin{equation}\label{sldexplicit}
                    \hat L_{\theta} =\sum_k \frac{\partial_\theta p_k}{p_k} |k\rangle\langle k|+2 \sum_{k,k'} \frac{p_k - p_{k'}}{p_k + p_{k'}} \,|k\rangle\langle\partial_\theta k| k' \rangle\langle k'|.
                \end{equation}
            \paragraph{Unitary transformations.} \label{qfiuntransf}
                So far we have kept the $\theta$-dependent transformation as much general as possible, getting cumbersome expressions. We now see that the restriction to a family of transformations (though very large) simplifies the expression for the Quantum Fisher Information to a large extent.\\
                We now focus on unitary transformations
                \begin{equation}
                     \rho(\theta) = e^{-i \theta \hat G} \rho_0 \,e^{+i \theta \hat G},
                \end{equation}
                where $\hat G$ is an Hermitian operator, the generator of the transformation.\\
                Inserting the identity operator $\text{Id}=e^{+i \theta \hat G}e^{-i \theta \hat G} $ in the definition of the SLD \ref{defsld}, we see that $\hat L_\theta = e^{-i \theta \hat G} \hat L_0 \,e^{i \theta \hat G}$, allowing us to rewrite it as 
                \begin{equation}\label{sldcommanti}
                    \{ \rho_0, \hat L_0 \} = 2i[\rho_0, \hat G].
                \end{equation}
                Considering \eqref{sldexplicit}, we can easily see that $\text{Tr}[\rho_0\hat L_0]=0 $ because $\hat L_0$ contains $p_k-p_{k'}$ and the trace introduces a $\delta_{k,k'}$, so
                \begin{equation}
                    F\big[ \rho_0, \hat G \big] = (\Delta \hat L_0)^2.
                \end{equation}
                Thus, the hardest part is to find the SLD from Eq. \eqref{sldcommanti}.\\
                For a pure state the QFI reaches its simplest form. In fact, considering \eqref{qfipurstat}, one obtains
                \begin{equation}
                \begin{array}{rl}
                    F\big[ \vert \psi (\theta) \rangle \big]&= 4 \big( \langle\partial_\theta \psi|\partial_\theta \psi\rangle - \vert \langle\partial_\theta \psi|\psi \rangle \vert^2 \big)\\
                    &=4\big(\langle \psi|\,i\hat G (-i\hat G) |\psi\rangle - \vert \langle \psi|\,i\hat G|\psi \rangle \vert^2\big)\\
                    &=4\big(\Delta \hat G\big)^2
                \end{array}.
                \end{equation}
                Then, computing the variance of the generator of the unitary transformation is enough to find the Quantum Fisher Information for a pure system state.
                
\subsection{Metrology implementations}
The Quantum Fisher Information provides a useful hint on the metrological usability of an entangled quantum state, however, it is not sufficient to devise accurate estimations of a quantum variable. In this section we recall the main concepts useful to frame this crucial question, and comment on their application in our specific case.\\   
First of all, the QFI is an upper bound for the Classical Fisher Information (CFI), obtained by its maximization over all possible quantum measurements. In fact, the proof of the Quantum Cram{\'e}r-Rao bound leads to the saturation condition $( \mathrm{Id} - \lambda_{\theta, \epsilon} \hat L_{\theta}) \hat{E}(\epsilon) = 0$.  Here, again, $\hat{E}(\epsilon)$ is an element of the Positive Operator Valued Measurements (POVM) set of projectors, which can be considered as a general kind of quantum measurements. Following~\cite{pezzesmerzipest:TinoKasevich}, it is evident that since the SLD $\hat L_{\theta}$ is an Hermitian operator, a set of POVM that satisfies the equality is represented by the projectors on the eigenstates of $\hat L_{\theta}$. This leads to the formal procedure used to find the optimal measurement providing the Classical Fisher Information (CFI) equal to the QFI.\\
Three problems hinder this procedure:
\begin{itemize}
    \item Finding the SLD is usually a difficult task, as one should know the density matrix of the system and solve the differential equation \eqref{defsld}.
    \item If one managed to find the SLD, it would not be certain of what kind of measurement one must perform to project on the SLD eigenbasis. In principle, it could be a difficult or even an impossible measurement.
    \item If the measurement on the eigenbasis of the SLD were experimentally viable, the SLD would still be dependent on $\theta$, the phase we want to measure. This implies that the choice of measurement depends on the quantity we need to measure. This can actually be addressed by an adaptive measurement scheme as described in~\cite{Barndorff}.
\end{itemize}
Here is where it becomes important to conceive protocols that ensure optimality. Besides optimality, one has to assess the robustness of the protocol with respect to experimental uncertainty, in particular the uncertainty on the number of particles. Work along these lines has been devoted in the context of quantum spin squeezing~\cite{Nolan,Smerzi_Loschmidt}, characterized by Gaussian Wigner distributions on the Bloch sphere~\cite{manori,Strobel}, and allow a relatively simple description of interferometric procedures in terms of rotations of vectors.\\
As stated in the main text, our ground states in the XY-FM and XY-FM Cluster phases might be represented by a Wigner function that is nonzero only in the neighborhood of the equator of the Bloch sphere. Indeed, as confirmed in our simulations, the expectation values of each component of the global spin operator, $\langle S_{x}\rangle=\langle S_{y}\rangle=\langle S_{z}\rangle=0$ are solid zero. Therefore, the Wineland spin squeezing parameter $\xi_R^2\equiv N\langle\Delta S_{\vec{n}_\perp}^2\rangle/|\langle\vec{S}\rangle|^2$\cite{wineland}, with $\vec{n}_\perp$ the direction perpendicular to the mean-spin direction $\langle\vec{S}\rangle/|\langle\vec{S}\rangle|$, cannot be defined. In addition, 
each system phase is characterized by small (but non zero) expectation values of $S_z^2$ as compared to $S_x^2\equiv S_y^2$, and the off-diagonal components of the spin covariance matrix (as e.g. $\langle S_x S_z\rangle$) are vanishing. As a result, the system is symmetric with respect to rotations around the spin z-axis, as also dictated by Hamiltonian (1) in the main text. Overall, from the simulation clues we may infer a description in terms of a ring-like Wigner function on the equator of the Bloch sphere.\\
As stated in the main text, this situation is compatible with the one considered in the experimental protocol realized by L{\"u}cke et al. \cite{Lucke} for Twin-Fock states. In the Bloch sphere representation, the protocol essentially amounts to tilt the ring-like state around an axis lying on the equatorial plane, and chosen so to minimize technical noise. From the perspective of the z-component, this implies an increase in variance of the probability distribution describing the outcomes. In the protocol, the information is encoded in the second moment of the z-component of the pseudo-spin, instead than in its first moment, i.e. the  population imbalance between the two spin states. After all, this would be consistent with the generalization of the QFI concept to many-body correlation functions expressed by Hauke et al. \cite{ZollerQFI}. Following~\cite{Lucke} and inspired from Kim et al.~\cite{Kim}, one can then write a lower bound for the CFI as $$\mathcal{F}\geq\frac{|d\langle S_z^2\rangle/ d\theta|^2}{(\Delta S_z^2)^2},$$ with $(\Delta S_z^2)^2\equiv\langle S_z^4\rangle-(\langle S_z^2\rangle)^2$. This would lead to an uncertainty $\Delta\theta\geq (\sqrt{\mathcal{F} n)}$, with $n$ the number of measurements, on the phase estimation.

\section{Spanning the phase diagram at quarter filling}
\subsection{Remark on filling and $J/U$}
As stated in the main text, $J/U$ is fixed so to mimic the on-site and nearest-neighbor parts of a same discretized interaction potential. For this purpose only, $J$ would better be small with respect to $U$. As to the sign, we remark that the original interaction term in the Hamiltonian (see main text) is taken to be of the form $V(|x_i-x_j|) c^\dagger_{i\uparrow}c_{i\downarrow}c^\dagger_{j\downarrow}c_{j\uparrow}$, so that discretization brings a minus sign via operators anticommutation. 

As to the filling parameter $\nu=1/4$, in the main text this was chosen as a compromise between a low-density system like in typical setups with ultracold fermionic gases, but not too low in order for the effects to be observable. Since the discussion is focused on the relation between the Quantum Fisher Information and the quantum phase diagram, we here report the details that are most relevant to understand the QFI behavior only for quarter filling. The effects due to larger commensurate, $\nu=1/2$, filling are illustrated in the main text.

\subsection{Correlations and phase diagram}
Here we show the analysis of the correlation functions, defined by $\langle O^\dagger_i O_j \rangle$, with $O_i$ being an operator acting on site $i$. Here is the list of the correlations analyzed in the present work:
\begin{itemize}
\item Spin $\alpha$ - spin $\alpha$ correlations: $O_i=s^\alpha_i$. As the system is invariant under rotations in the $x,y$ spin plane, we will analyze only $x$ and $z$ components, having checked that the $y$ component behaves exactly like the $x$ one.
\item Density - density correlations: $O_i=n_i$
\item Pair - pair correlations: $O_i=c_{i\uparrow}c_{i\downarrow}$
\end{itemize}
These correlations allow us to observe the typical behaviors of a 1D fermionic quantum system, such as Charge Density Waves, Spin Density Waves and SuperFluid-like ordering. The information contained in these quantum correlations, together with the ground state expectation values of on-site operators, can shine light on the nature of the system's phases~\cite{giamarchi}.\\
\subsection{Averaging procedure}
In order to represent these correlations in a proper manner, we must consider the presence of Friedel oscillations~\cite{friedelosc}. Their wavelength is constant and equal to $1/(2k_F)$, so we rule out their effects via the following averaging procedure. Our algorithm returns the correlations between any pair of sites, so we start from a point far from the edges (for $L=100$ we consider e.g. site 20 from the left edge to be far enough) and take the correlations between this point and all the others away from the edges (e.g. 50 points to its right). We then take the correlations between the same amount of sites and the next point (21 in this example) and sum them to the previous ones, site by site. We repeat this procedure for a number of neighboring starting points, that be proportional to the Friedel oscillation's wavelength. Finally, we divide the result by the number of considered points. This procedure eliminates almost completely the effect that  the starting-point position has on the correlation function, without modifying the $2k_F$ component of the observed correlations. This is crucial for SDW and CDW orderings. This is due to the fact that all the oscillating correlations \textit{share the same phase with respect to the starting point}. This means that we can safely average them once we sum the correlations at the same distances.
\begin{figure}
\includegraphics[width=0.95\columnwidth]
{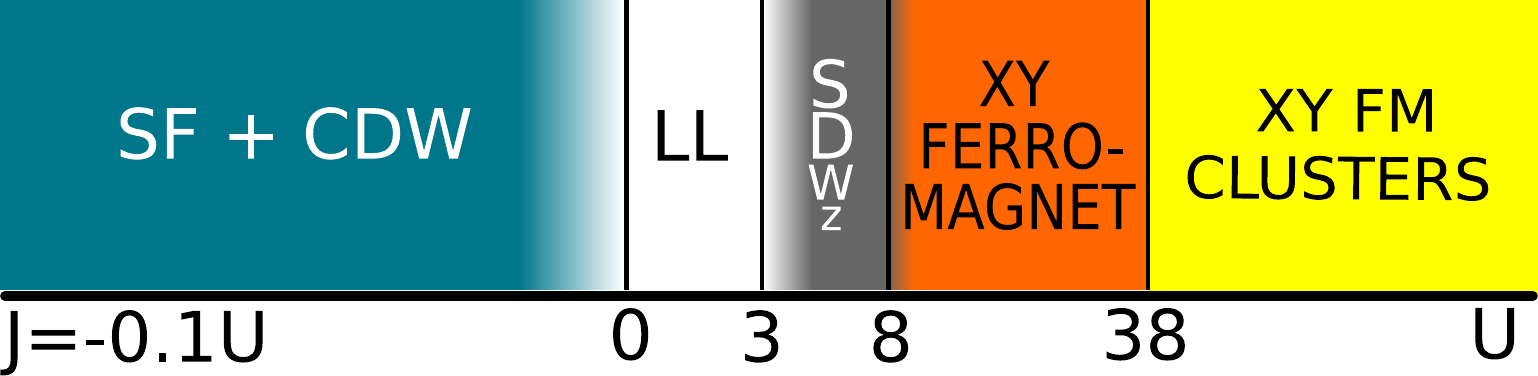}
\caption{Quantum phase diagram for $\nu=1/4$ and $J=-0.1U$. This is a single line in the global phase diagram depicted in the main text in Figure 1.}
\label{smfig3}
\end{figure}
\begin{figure}
\includegraphics[width=0.95\columnwidth]
{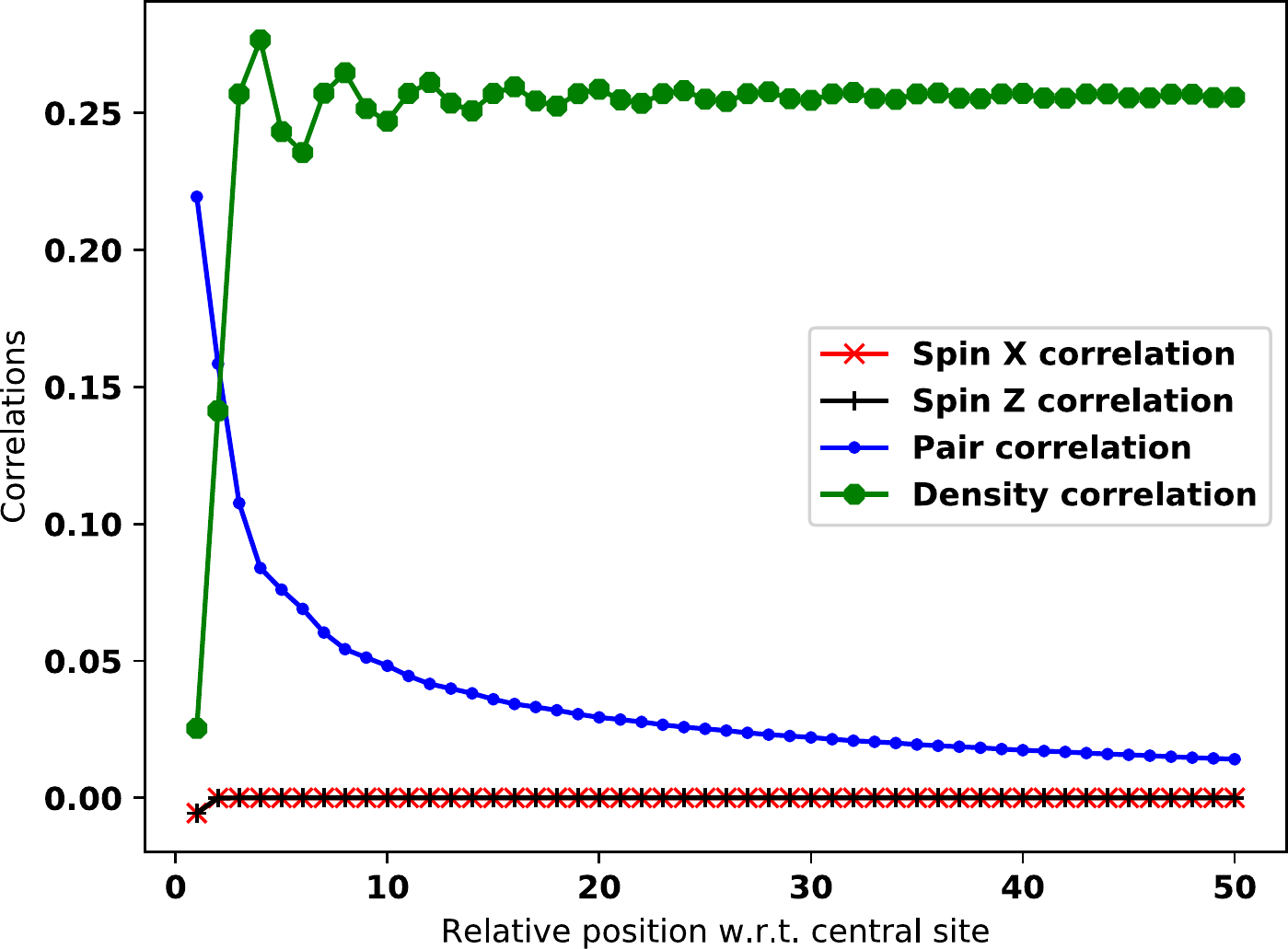}
\includegraphics[width=0.95\columnwidth]
{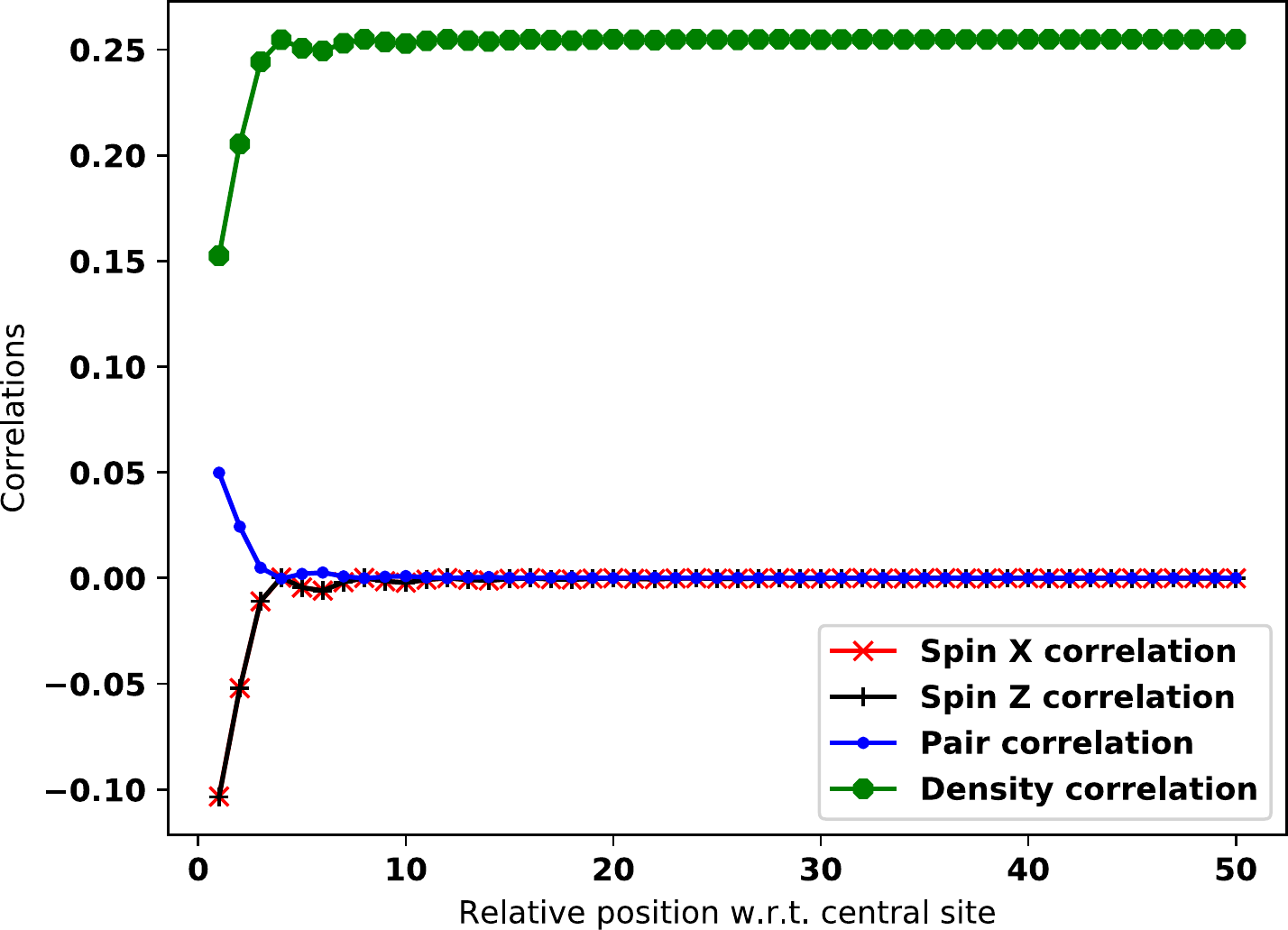}
\includegraphics[width=0.95\columnwidth]
{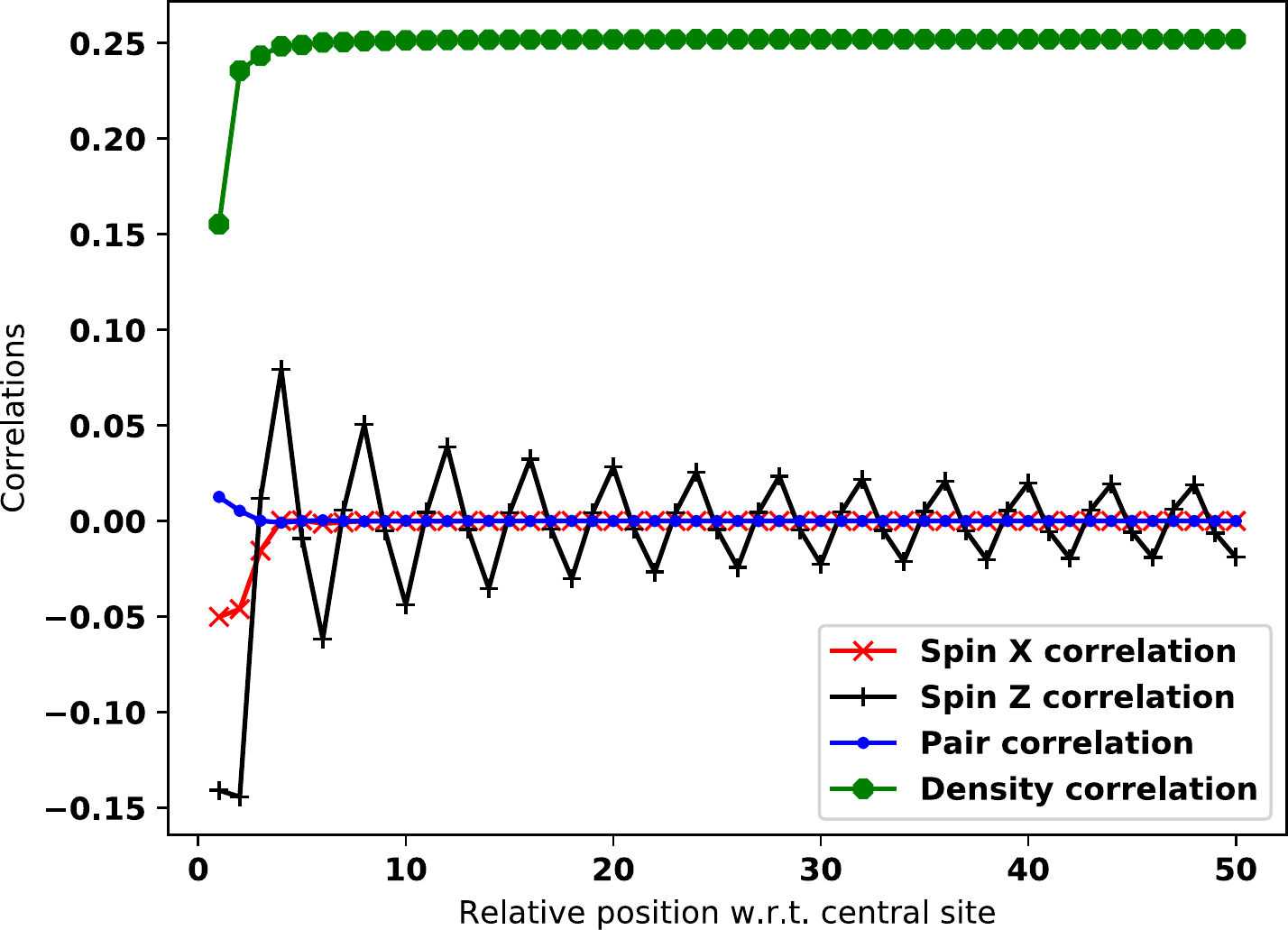}
\caption{Correlations and quantum phases. Phases of the $tUJ$ model at quarter filling (one particle every two sites) and $J=-0.1U$. The three phases with a small QFI. Top: the SF for $U=-20$, where a nonzero CDW signature is also found. Middle: LL phase for $U=0.1$. In this phase, each correlation has a small but non-vanishing expectation value. Bottom: SDW$^z$, while the system shifts from the LL to the next coming XY-FM phase (see below). Here, $L=100$ and $m=300$.}
\label{smfig4}
\end{figure}
This being said, we can begin the analysis of the correlation functions. 
\subsection{Correlations and quantum phases}
For the purposes of this work, we focus here on a qualitative analysis of the correlation functions, that is sufficient to identify the different phases. Quantitative analysis performed via a fitting procedure confirms the qualitative findings.
In the following, the simulation parameters are $L=100$ and $m=300$. Where not stated otherwise, the expectation values for the on-site operators correspond to the following:
\begin{itemize}
    \item Density operator: constant (0.5) with superimposed Friedel oscillations
    \item Spin operators: all zero on any site.
\end{itemize}

The phases can be analyzed both in general and from the QFI point of view. We remind from the main text that the Quantum Fisher Information is computed as the biggest sum of all the spin correlations among all spin components. In our system, the most important correlation for QFI always appears to be the spin-$x$ component. In the following, we cross all the different phases encountered in the main text. As the first five are encountered in the illustrative example $J/U=-0.1$, we use just the $U$ value to identify their position in the phase diagram, referring to Fig.\ref{smfig3} for a graphical representation. 
\subsubsection{SuperFluid (SF)}
At $U=-20$ we observe that most fermions are in a doubly occupied state. This is confirmed by the correlation function plotted in the top panel of Fig.~\ref{smfig4}, where we can see the dominance of the pair correlation on the others and the presence of a CDW ordering as well. The formation of pairs is driven by the attractive effect of the negative on-site coupling $U$.\\
This phase carries a vanishing QFI. We can find a reason for it in pair formation. By definition, pairs are spin singlets. Therefore, they belong to a $j=0$ representation of angular momentum, that prevents them to have any spin ordering in any direction.
\subsubsection{Luttinger Liquid (LL)}
In the vicinity of $U=0$, we can observe the Luttinger Liquid (LL) phase. This phase is characterized by the presence of a non-vanishing expectation value for all correlations, as displayed in the middle panel of Fig.~\ref{smfig4}. The position of the LL in the phase diagram for $U\sim J\sim0$, agrees with expectations ~\cite{giamarchi,sciampa1,sciampa3}.\\
This phase has a vanishing QFI, as spin correlations are SDW-like and small.
\subsubsection{Spin Density Wave-$z$ (SDW$^z$)} 
While increasing $U$, we observe a quite sudden rise of the SDW ordering along the $z$ component accompanied by almost zero SDW$^{x,y}$ correlations, as depicted in the bottom panel of Fig.\ref{smfig4}. \\
This phase has a vanishing QFI too, because the only non-zero spin correlation is oscillating, implying a zero sum.
\begin{figure}
\includegraphics[width=0.95\columnwidth]
{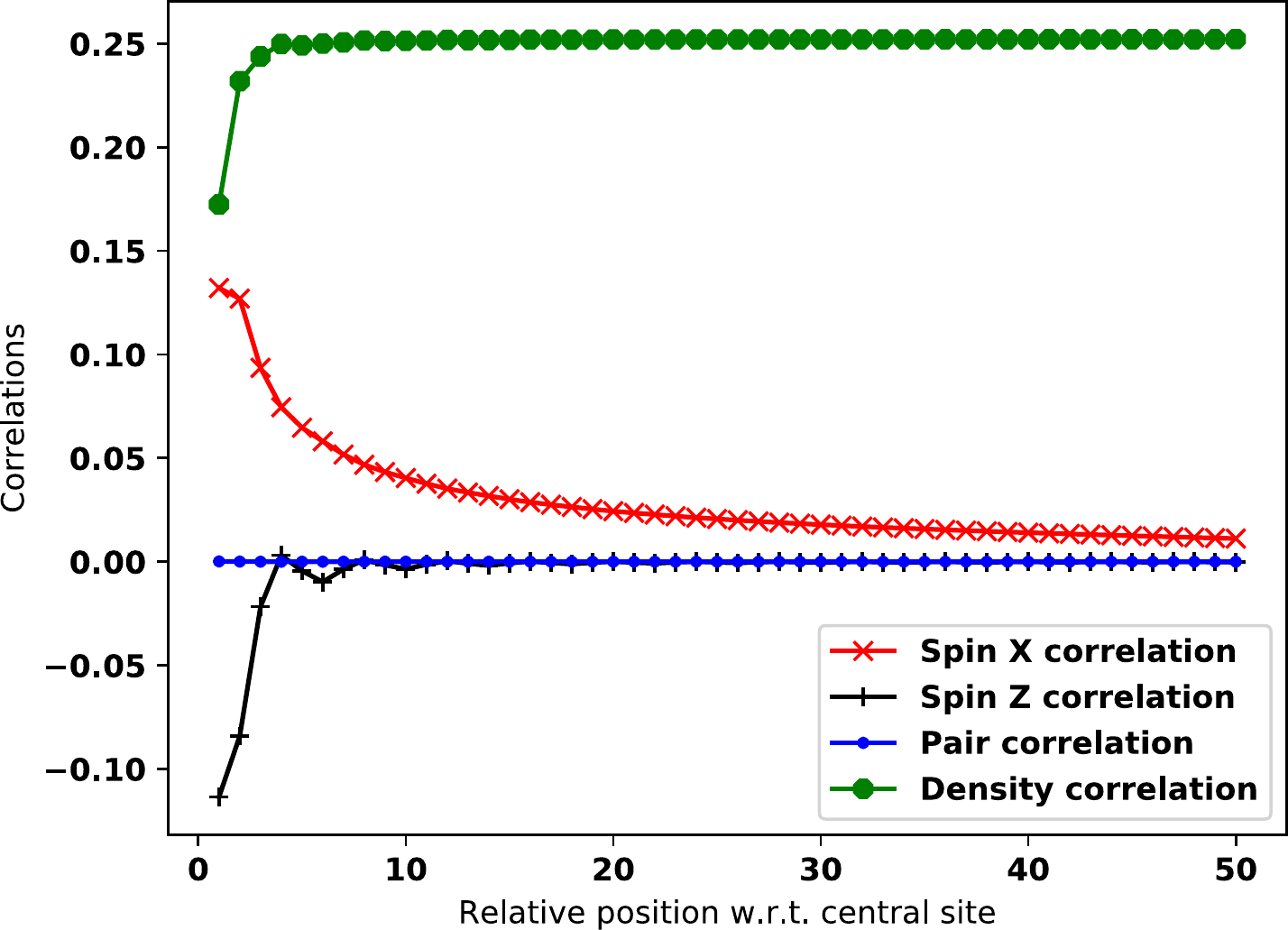}
\includegraphics[width=0.95\columnwidth]
{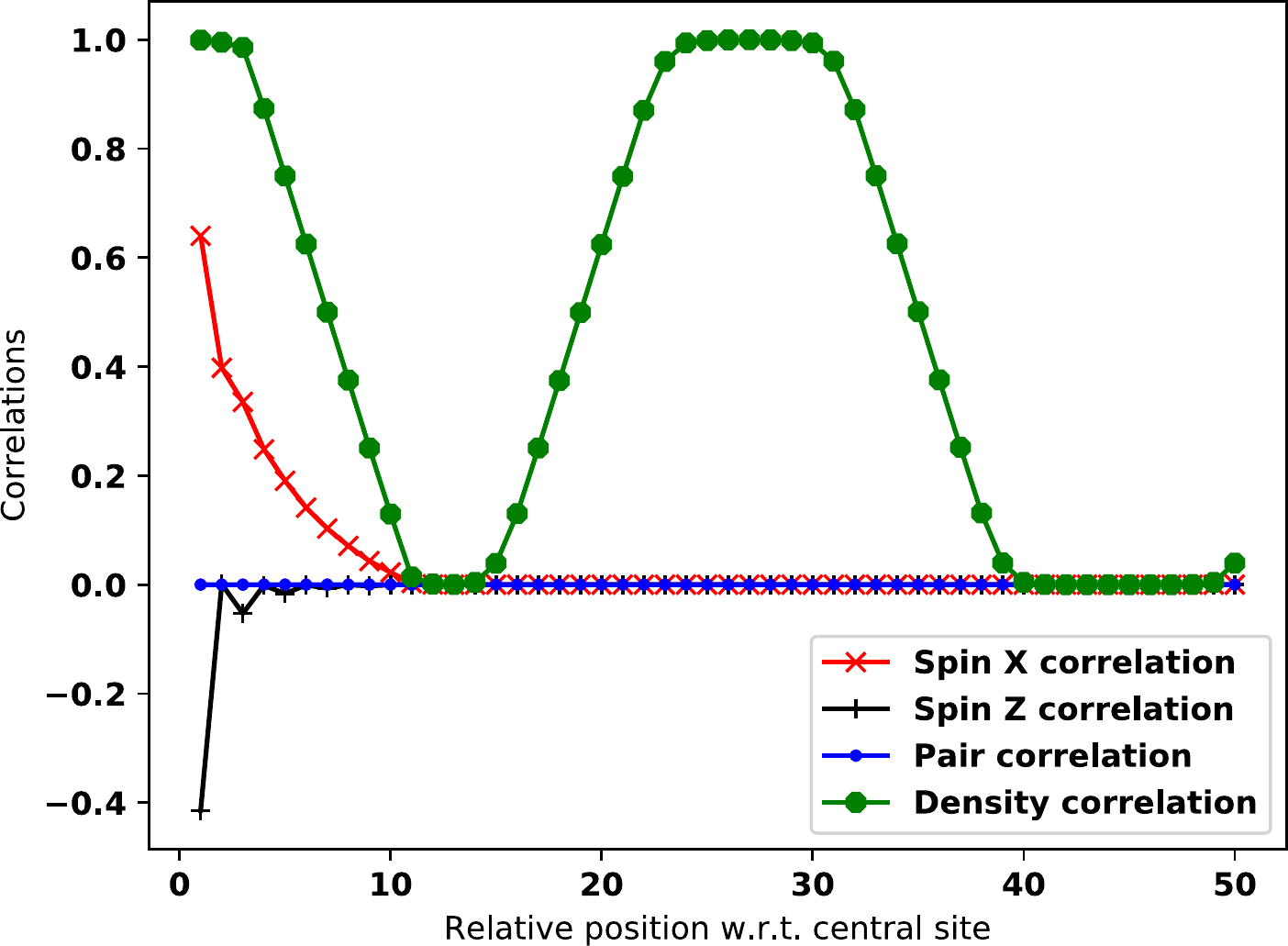}
\caption{Correlations and quantum phases. Phases of the $tUJ$ model at quarter filling and $J=-0.1U$. The two phases with large QFI. Top: XY-FM phase, characterized by non-oscillating decaying correlations in the spin-$x$ component, with positive values. Bottom: XY-FM Cluster phase, characterized by  strong correlations in the spin $x,y$ plane,  confined in a part of space. Here, $L=100$, $m=300$.}
\label{smfig5}
\end{figure}
\subsubsection{XY-FerroMagnet (XY-FM)} 
When $U\sim6$ at quarter filling, the system quickly shifts to a phase with dominating positive correlations in the spin $x,y$ plane. CDW, SF pairs and SDW$^z$ are absent in this regime, as shown in the top panel of Fig.\ref{smfig5}.\\
These spin-$x,y$ correlations sum up to a non-zero value, implying a large Quantum Fisher Information (see the main text).
\subsubsection{XY-FerroMagnetic Clusters (XY-FM Clusters)}
This phase is characterized by clustering in strongly spin-correlated droplets. In this regime , the spin-$x$ correlation function has positive and large values, though confined inside each cluster as shown at the bottom of Fig.\ref{smfig5}. This means that particles belonging to two different clusters are completely spin-uncorrelated.
Strong spin correlations should induce a large QFI  value, but uncorrelation between different clusters actually lowers it (see main text).

\subsubsection{Hemmed Clusters (HC)}
\begin{figure}[h]
\includegraphics[width=0.95\columnwidth]{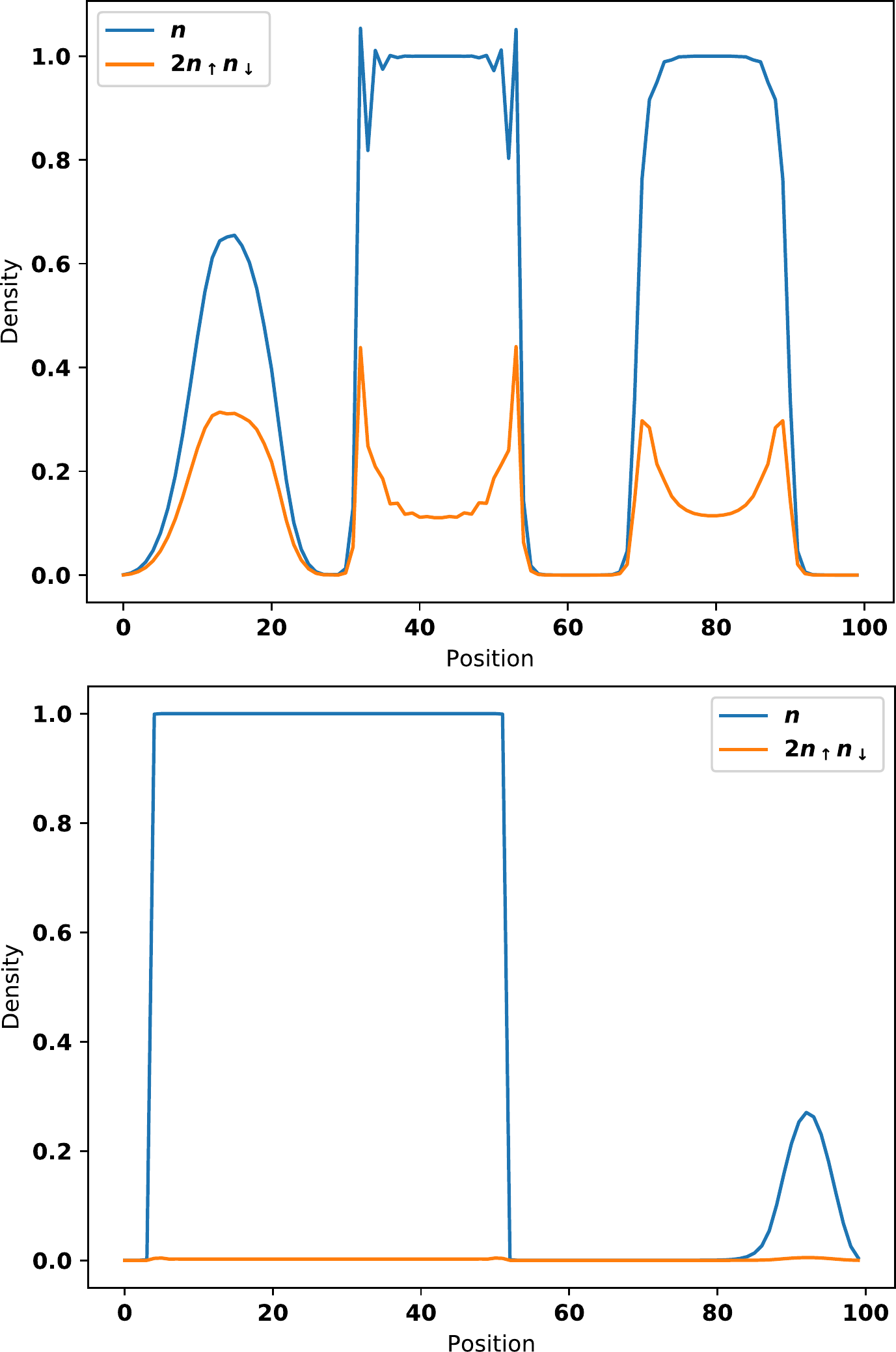}
\caption{Correlations and quantum phases at quarter filling. Hemmed clusters (HC). Top: overall density and double-occupation density in the HC phase at $J/U=-0.9$ and $U=-14$. The HC phase appears for $J\sim-0.9U$ at $U\sim -15$. The HC edges contain a large fraction of doubly occupied sites, that seem to be more stable than the normal "Heisenberg spin-chain" edges in this regime. This may occur because $U$ is larger than $J$: while $J$ is positive, the spin-ordering and pair formation mechanisms leading to energy-lowering may take place at the same time. 
Bottom: density plot for $J/U=-2$ and $U=-20$ deep in the XY-AFM cluster phase. Compared to the top panel, double occupancy shows no appreciable spikes at the cluster edges.}
\label{smfig6}
\end{figure}
In this phase cluster formation is observed  with mainly XY-AntiFerroMagnetic (XY-AFM) ordering. The peculiarity of this phase is represented by a non-negligible fraction of particles in a doubly occupied state. 
As an intuitive explanation, we could argue that the system organizes the cluster edges in pairs because the pair is energetically more convenient than a spin-chain loose end. Edges cannot completely be in a paired state, otherwise this would once more create loose ends in the spin chain. \\
An amusing analogy can be found with clothing: as the edge of a piece of cloth can become unweaved, a skillful tailor folds the edges on themselves, creating a hem. In the absence of pairs, the spin chain would be longer but weaker. The double occupancy can then be seen as the loose end folded back on itself, so it becomes a kind of "hem". The corresponding density profile is pictured in top Fig.\ref{smfig6}, and is to be contrasted with the bottom profile for our conventional clusters.
\subsubsection{Charge Density Wave (CDW)}
\begin{figure}
\includegraphics[width=0.95\columnwidth]
{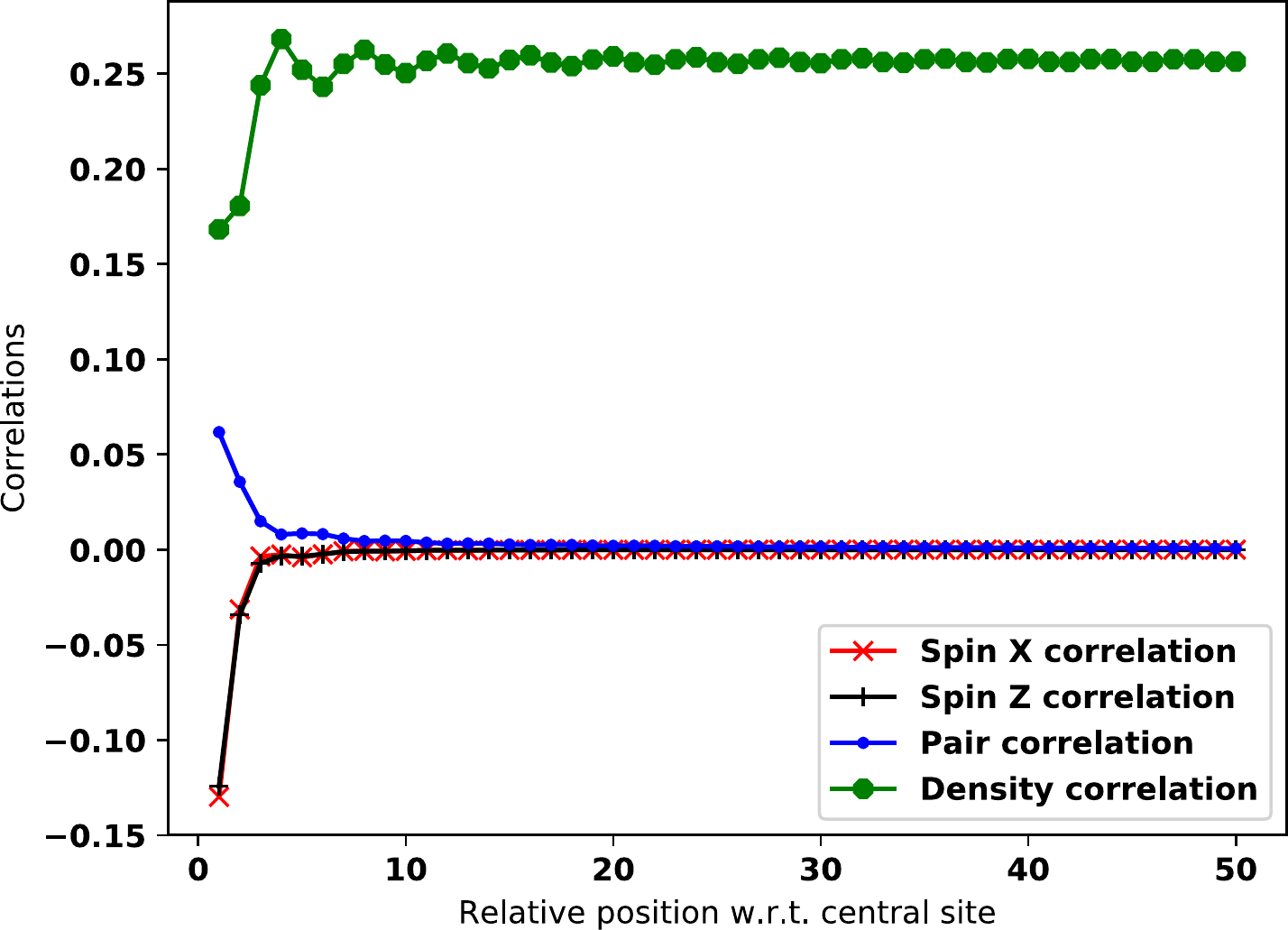}
\includegraphics[width=0.95\columnwidth]
{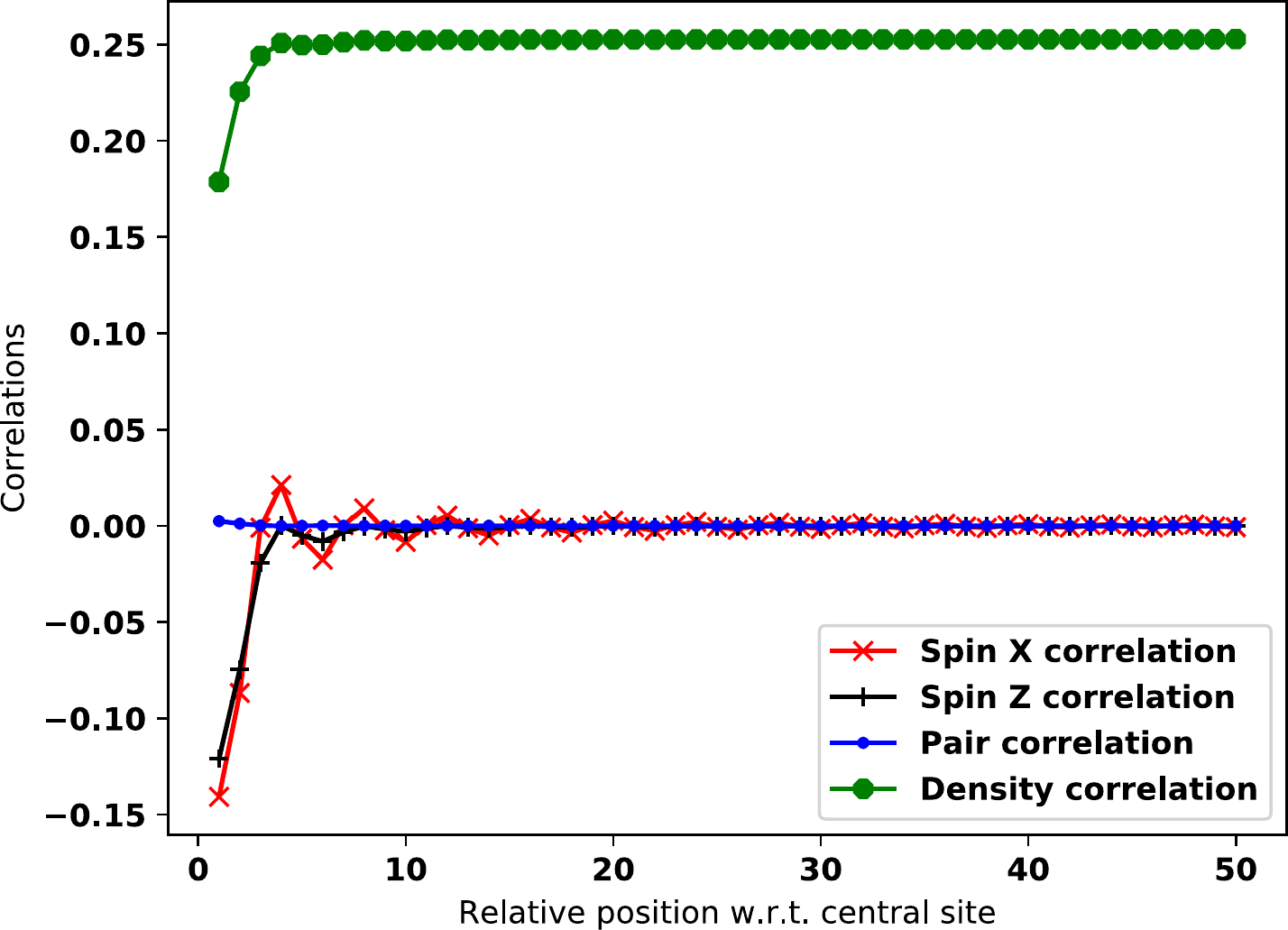}
\includegraphics[width=0.95\columnwidth]
{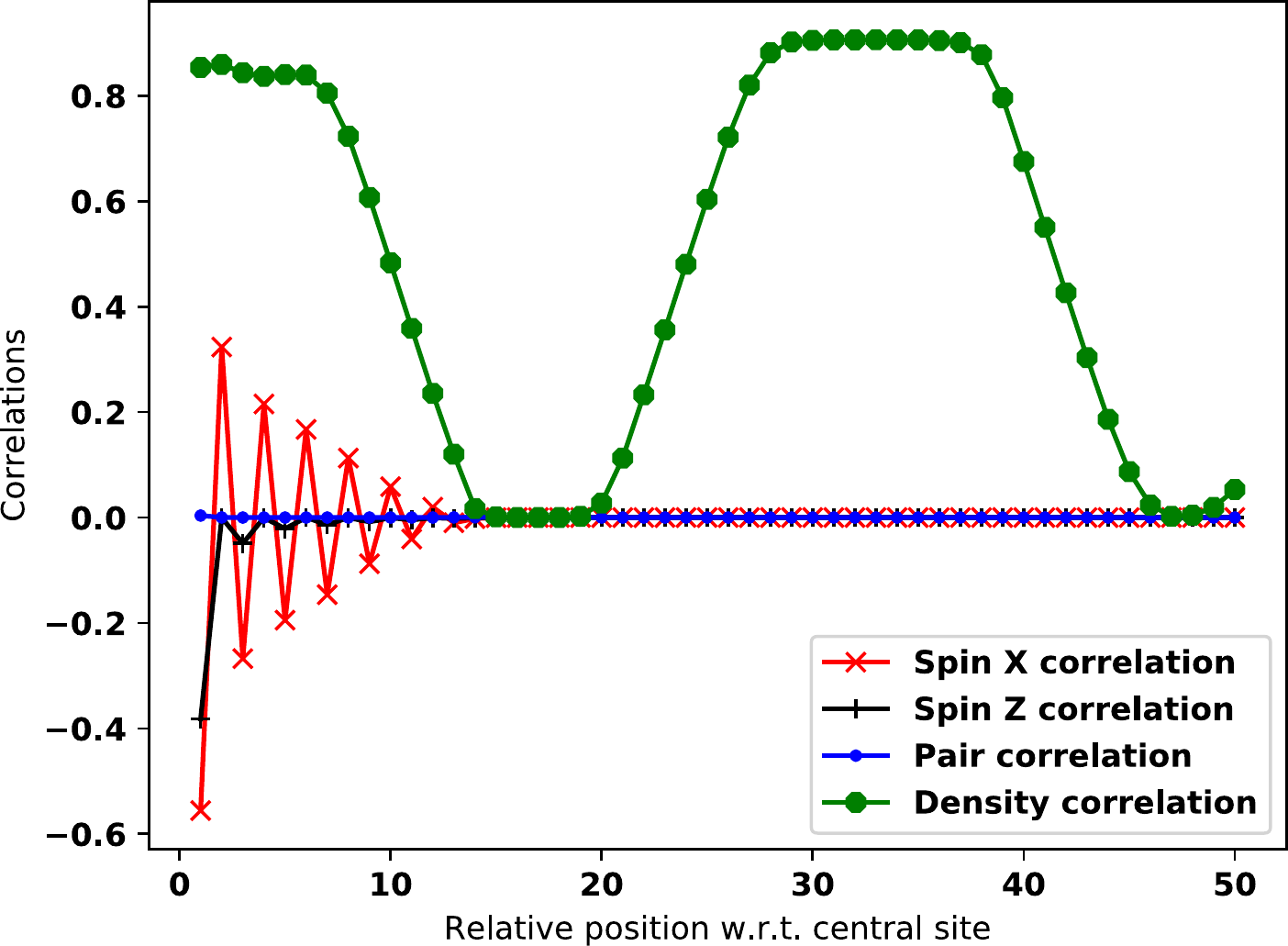}
\caption{Correlations and quantum phases at quarter filling. Phases of $tUJ$ model. Top: correlations $J=-1.5U$, $U=-0.8$ in the CDW phase, manifesting with an oscillatory behavior at $1/(2k_F)$ wavelength. Even though not pronounced, its existence is more visible when compared to the same plot in the middle panel, where there is no CDW ordering.
Middle: correlations at $J=0.07U$, $U=15$ in the SDW$^{x,y}$, manifesting with an enhanced oscillatory behavior at $1/(2k_F)$.  
Bottom: correlations at $J=0.8U$, $U=8$ in the cluster phase at $J>0$. Large antiferromagnetic (AFM) correlations are visible. Noticing that the density correlations get the form of a density profile, the AFM correlations can be considered to build up in the cluster interior.}
\label{smfig7}
\end{figure}

In this phase, we observe a decaying oscillation in the density-density correlation function, displayed on top Fig.~\ref{smfig7}. CDW ordering is present as well in the SF phase, but it becomes dominant in the regime of small $|U|\sim0$ and $1\lesssim J\lesssim 4.5$. We observe it also for $U=0$, so this can be considered as a pure $J$-driven ordering such as that leading to cluster formation. This ground state configuration has no effect on the QFI, as only spin orderings can modify it.
\subsubsection{Spin Density Wave-$x,y$ (SDW$^{x,y}$)}
This phase is obtained by setting the system parameter to large $U>0$ and small $J>0$. It is characterized by the presence of a Spin Density Wave in the spin-$x,y$ component. While the other phases close to LL appear for relatively small differences in the coupling values from the free case with $U=J=0$, i.e. for $|U|,|J|\sim1$, this phase occurs much farther from the origin ($U\sim4$,$J\sim0.3$). The corresponding correlations are pictured in middle Fig.~\ref{smfig7}, manifesting as decaying oscillations at $1/(2k_F)$ wavelength. 
\subsubsection{XY-AntiFerroMagnetic Clusters (XY-AFM Clusters)}
For large and positive $J\gtrsim5$ and a wide range of $U$ values, we observe a second cluster phase. Unlike the XY-FM Cluster case, positive $J$ values favor anti-alignment of the spin-$x,y$ components. This AFM behavior is exactly what we observe in this configuration, that shares with its ferromagnetic counterpart the uncorrelation behavior between different clusters (at quarter filling). As displayed in bottom Fig.~\ref{smfig7}, in the XY-AFM Clusters the correlation sign changes every each site. This behavior is different with respect to that characterizing SDW ordering, where the sign changes with a $1/(2k_F)$ period \cite{giamarchi}.

\subsection{Dependence on filling and on statistics} 
In order to assess the degree of generality of our conclusions, we have performed simulations with additional values of the filling factors as well as for the case of particles with bosonic statistics.\\
In particular, we have changed the commutation rules to bosonic on different sites, while retaining the maximum double occupancy on the same site. For the cluster and superfluid phases we have observed similar results with respect to the fermion case, with compatible QFI scalings. This observation leads us to infer that whenever clusters form, the (anti-)symmetrization of the many-body wavefunction be confined in the motional degrees of freedom.\\
We have then performed simulations for filling factors different from the $\nu=1/4,1/2$, in particular $\nu=1/3$ and $n/20$ with $n=3$, 7, 9, 11. In fact, from a preliminar analysis, a non-trivial scattered scaling behavior of the QFI emerges, possibly caused by frustration. This requires better suited investigation tools and more extended studies, which are referred to future work~\cite{Lepori}.\\


\end{document}